\documentclass[journal,10pt]{IEEEtran}
\IEEEoverridecommandlockouts
\usepackage{graphicx}
\usepackage{caption}
\usepackage{enumerate}
\usepackage{epsfig}
\usepackage{epstopdf}
\usepackage{svg}
\usepackage{cite}
\usepackage{subfigure}
\usepackage{amsmath,amssymb,amsfonts}
\usepackage{algorithmic}
\usepackage{tabularx}
\usepackage{textcomp}
\usepackage{xcolor}
\usepackage{stfloats}
\usepackage{multirow}
\usepackage{url}
\usepackage[ruled]{algorithm2e}
\usepackage{hyperref}
\usepackage{etoolbox,xstring,mfirstuc,textcase}
\usepackage{bm}
\usepackage{amsmath}
\usepackage{amsthm}
\usepackage{stackengine}
\usepackage{cuted}
\usepackage{makecell}
\usepackage{booktabs}
\usepackage{threeparttable}
\usepackage[skip=3pt]{caption}
\usepackage[explicit]{titlesec}

\newtheorem{lemma}{Lemma}
\newtheorem{remark}{Remark}
\def\BibTeX{{\rm B\kern-.05em{\sc i\kern-.025em b}\kern-.08em
    T\kern-.1667em\lower.7ex\hbox{E}\kern-.125emX}}

\titlespacing*{\section}{0pt}{6pt}{0.5pt}  
\titlespacing*{\subsection} {0pt}{4pt}{0.5pt}
\begin{document}

\title{Stacked Intelligent Metasurface-Enhanced Wideband Multiuser MIMO OFDM-IM Communications} 
	\author{Zheao Li, \IEEEmembership{Graduate Student Member, IEEE}, Jiancheng An, \IEEEmembership{Senior Member, IEEE}, and Chau Yuen, \IEEEmembership{Fellow, IEEE}
\vspace{-0.8 cm}

\thanks{Reference: {\color{blue}{Z. Li, J. An and C. Yuen, “Stacked Intelligent Metasurface-Enhanced Wideband Multiuser MIMO OFDM-IM Communications," \emph{IEEE Trans. Wireless Commun.}, 2026, doi: 10.1109/TWC.2026.3713490}}. An earlier version of this paper was presented in part at the IEEE GLOBECOM 2025~\cite{GC-SIM}. (\emph{Corresponding author: Chau Yuen}.)}
\thanks{Z. Li, J. An, and C. Yuen are with the School of Electrical and Electronic Engineering, Nanyang Technological University, 639798, Singapore (email: zheao001@e.ntu.edu.sg, jiancheng.an@uestc.edu.cn, chau.yuen@ntu.edu.sg).}
}

\markboth{Accepted to IEEE TRANSACTIONS ON WIRELESS COMMUNICATIONS, doi: 10.1109/TWC.2026.3713490.}%
{Shell \MakeLowercase{\textit{et al.}}: A Sample Article Using IEEEtran.cls for IEEE Journals}

\maketitle

\begin{abstract}
Stacked intelligent metasurfaces (SIM) enable fine-grained wave-domain signal processing, but their wideband deployment is impeded by two structural factors: (i) a single, quasi-static SIM phase tensor must adapt to all subcarriers, and (ii) multiuser scheduling changes the subcarrier activation pattern frame by frame, requiring rapid reconfiguration. To address these, we propose a SIM-enhanced wideband multiuser transceiver built on orthogonal frequency-division multiplexing with index modulation (OFDM–IM). The sparse activation of OFDM-IM confines high-fidelity equalization to the active tones, effectively widening the usable bandwidth. To make the design reliability-aware, we directly target the worst-link bit-error rate (BER) and adopt a max-min per-tone signal-to-interference-plus-noise ratio (SINR) as a principled surrogate, turning the reliability optimization tractable. For frame-rate inference and interpretability, we propose an unfolding projected-gradient-descent network (UPGD-Net) that unrolls across the SIM’s layers and algorithmic iterations with a learnable per-iteration step size. Simulations demonstrate that the proposed framework achieves fast convergence and significant BER gains over fully-digital baselines. Notably, the design exhibits superior robustness against errors and outperforms large-aperture hybrid precoding benchmarks in both sum rate and energy efficiency. By combining structural sparsity with a BER-driven, deep-unfolded optimization backbone, the proposed framework effectively resolves the key wideband deficiencies of SIM.

\end{abstract}
\section{Introduction}\label{sec:intro}

\IEEEPARstart{P}{rogrammable} metasurfaces have revolutionized the manipulation of electromagnetic (EM) fields and waves, which play a significant role in future sixth-generation (6G) wireless systems~\cite{PM,PM1,PM4,6GMIMO}. Composed of densely packed, sub-wavelength meta-atoms with tunable responses, advanced metasurfaces can be employed to support various kinds of wireless communication functions~\cite{RIS, PM2, PM3,PM5}. Among these, stacked intelligent metasurfaces (SIM) have emerged as a fully–analog means of executing {wave-based computing} directly in the EM wave domain for multiple-input multiple-output (MIMO) beamforming~\cite{SIMHMIMO}. By cascading several metasurface layers along the propagation direction, a three-dimensional SIM can realize tunable signal processing and gain computing ability in the wave domain: each layer imposes an adjustable phase-shift profile on the impinging wavefront, and the cascade shapes the spatial transmission function \cite{SIM1,SIM2,SIM3}. In practice, SIM can be equipped on the antenna arrays, functioning as an intelligent radome, offering greater flexibility in controlling EM wave propagation. This fully-analog architecture reduces the extra need for digital baseband processing, minimizes the number of radio frequency (RF) chains, and allows the use of low-resolution digital-to-analog converters (DAC) and analog-to-digital converters (ADC), resulting in lower hardware complexity and improved energy efficiency~\cite{SIM-EE}. Compared with fully-digital and hybrid arrays, SIM shifts multiuser beamforming out of baseband, bringing substantial hardware/energy savings while preserving fine wavefront control. Besides, SIM addresses this limitation by providing depth with multiple sheets rather than single-layer metasurfaces, thereby providing more degrees of freedom (DoF) than traditional reconfigurable intelligent surfaces (RISs) and holographic metasurface antennas~\cite{SIMHMIMO2,SIM-RIS}.

Recent studies have demonstrated SIM’s versatility across a range of applications. For instance, \cite{SIM-HMIMO, SIM-HMIMO-MI} illustrated SIM’s potential for highly controllable signal transmission and reception in holographic MIMO systems. SIM-based beamforming techniques were explored in \cite{SIM-BF2,SIM-BF3,SIM-BF5} to demonstrate their superiority over conventional MIMO in energy efficiency. Multiuser communication performance enhancements by SIMs were reported in \cite{SIM3,SIM-CE2,SIM-MultiUser1,SIM-MultiUser2,SIM-BF4,SIM-CE1}, emphasizing their adaptability and effectiveness. SIM has also been applied to direction-of-arrival (DoA) estimation~\cite{SIM-DOA}, satellite communication~\cite{SIM-LEO}, communication security~\cite{SIM-Secure}, and semantic communication~\cite{SIM-SC}. In both cellular and cell-free networks, SIM has shown promise in mitigating inter-user interference (IUI) and enhancing resource allocation~\cite{SIM-CF,SIM-CF2,SIM-dp}. Moreover, acting like an artificial neural network (ANN), SIM packs multiple layers of metasurfaces for advanced computation and signal processing tasks, with the configured transmission coefficients serving as trainable network parameters. Therefore, SIM can be employed to support many beyond-communication functionalities, such as wireless sensing and localization~\cite{SIM-HW, SIM-S2,SIM-S3}, while recent efforts have investigated reinforcement learning and multi-agent frameworks to dynamically configure SIM for various scenarios~\cite{DRL2,MetaLearning}. In summary, most previous works on SIM focus on narrowband systems with frequency-flat channels~\cite{SIMsur1, SIMsur2}, where analog signal processing can be easily tackled with high energy efficiency in the wave domain.

Extending SIM from narrowband prototypes to wideband orthogonal frequency-division multiplexing (OFDM) has shown clear promise. The multi-interaction of EM waves through SIM can be viewed as an analogue multi-path synthesizer, where the transmission via each meta-atom acts as a tap with adjustable channel impulse response (CIR) weight~\cite{Wideband-SIM2}. By refining the layer-wise phase shifts, the cascaded device can be configured to combat delay dispersion and enhance multi-carrier transmission. In fact, recent studies reported that fully–analog transceivers equipped with wideband SIMs improve spatial multiplexing and robustness against frequency-selective deep fades without resorting to digital precoders and combiners \cite{Wideband-SIM,Wideband-SIM2}. However, two structural bottlenecks persist. First, within an OFDM symbol the SIM phase profiles are quasi-static and hence common to all subcarriers, whereas the per-tone transfer factors vary with frequency. The resulting beamformer must compromise simultaneously across all subcarriers. With moderate aperture and layer count, the performance of SIM is closely tied to the physical and EM properties of the materials used, limiting the specific operational bandwidth. Second, in multiuser MIMO OFDM with dynamic scheduling, the subcarrier allocation changes every frame. In principle, each new subcarrier allocation strategy requires recomputing the entire SIM phase-shift profile under unit-modulus constraints. These facts motivate a transmission and optimization framework that \emph{relaxes the “equalize–all–tones’’ requirement imposed by cross-band coupling} while \emph{enabling frame–rate reconfiguration}.

OFDM with index modulation (OFDM–IM) provides the required structural complement for wideband SIM applications from the coding and modulation side. In each OFDM-IM subblock, only a subset of subcarriers is activated, and their positions, which are determined by the incoming index bits, serve as an additional information-bearing dimension~\cite{IM, IM1,IM2}. The induced spectral sparsity (i) lowers peak-to-average-power ratio (PAPR) by reducing the probability of large-symbol superposition in time~\cite{PAPR}, and (ii) requires high-quality equalization only on the active tones, thereby effectively widening the bandwidth over which a given SIM can realize fully–analog beamforming. {{Specifically, since the subcarrier activation set is a subset of the full spectrum, optimizing high-dimensional SIM phase-shift matrices in OFDM-IM is less constrained than in full-carrier systems. This effectively relaxes the wideband beamforming requirements, allowing the frequency-static SIM phases to achieve higher equalized gain on the targeted active tones.}} Prior work~\cite{GC-SIM} quantified this trade-off of activation pattern selections in a SIM-enhanced multiuser system, which formulated the channel fitting optimization problem with subcarrier allocation to relax the interference-mitigation constraints. In addition, RIS-aided OFDM-IM designs were proposed in~\cite{RIS-IM1, RIS-IM2} to leverage the subcarrier allocation to transmit extra information efficiently, where phase shifts were configured by an exhaustive search-based optimization algorithm. However, no SIM-enhanced multiuser OFDM–IM transceiver benchmarked against a bit error rate (BER) metric has been established. Bridging this gap advances the SIM designs from narrowband/rate-centric formulations to a wideband, reliability-aware, fully-analog architecture.

The joint design, however, leads to a highly coupled, frame-adaptive optimization: the highly-coupled SIM phase shifts, per-tone power, and the subcarrier activation matrix determined by the current index bits must be updated every OFDM frame under unit-modulus constraints. Classical semi-definite relaxation (SDR) or interior-point methods incur high polynomial complexity and must be restarted for each new subcarrier activation strategy, violating latency targets. The closed box deep neural networks (DNNs) are fast but often ignore physical constraints, yield infeasible phases, and typically have poor interpretability~\cite{DUN}. A principled alternative is the \emph{deep unfolding network} (DUN), which rewrites a finite number of algorithmic iterations as a transparent, layer-wise neural network architecture~\cite{DUN1,DUN2}. Constraints of optimization problems (e.g., unit–modulus projection) are enforced per layer, monotonic descent is preserved, and only a handful of hyper-parameters (e.g., step sizes) are learned. Because the subcarrier activation matrix enters explicitly in each unfolded stage, a single trained network naturally adapts to frame-varying activation patterns, which precisely correspond to the operating regime of OFDM-IM. Prior DUNs for downlink beamforming~\cite{DUN3, DUN4, DUN5} mainly target sum rate or signal-to-noise-ratio (SNR) proxies to simplify complex tasks with lower computational complexity; explicit BER-driven SIM optimization remains fully unexplored.

To fill the above research gaps, we develop a SIM-enhanced multiuser OFDM-IM transceiver and a physics-guided deep-unfolding solver, tailored to wideband communications. A multilayer SIM performs wave-domain zero-forcing (ZF) to mitigate IUI, while OFDM-IM supplies frequency-domain sparsity that lowers PAPR and relaxes the equalization burden. By grouping nearest-neighbor error events into distance classes, we establish a bridge from the worst-link BER to an active-link per-tone signal-to-interference-plus-noise-ratio (SINR). This converts BER minimization into a principled surrogate, formulated as a constrained max-min SINR optimization problem. The projected-gradient-descent (PGD) for this program is then unfolded into a coupled physical-algorithmic network, preserving full interpretability. Each cell executes a closed-form PGD update with unit-modulus projection, while only the per-iteration step sizes are learned offline to avoid overfitting. The training database can cover hundreds of channel/activation patterns.

The main contributions and innovations of this research are described as follows.
\begin{itemize}
\item We propose the first SIM-enhanced wideband multiuser MIMO OFDM-IM transceiver: a reconfigurable multi-layer SIM realizes fully-analog ZF precoding in the wave domain, and OFDM-IM supplies frequency-domain diversity. The joint design suppresses IUI, mitigates wideband frequency-selective fading without digital precoding, and improves PAPR profile and energy efficiency.
\item We derive an analytical BER-to-SINR bridge under SIM-enhanced beamforming and convert BER minimization into an active-link max-min SINR surrogate problem. The problem reformulation turns a non-convex BER objective into {{a tractable surrogate for index-aware phase optimization}}, closing the gap between SINR heuristics and true error-rate optimization.
\item {{We design a coupled physical-algorithmic unfolding PGD network (UPGD-Net) aligned with SIM physics:}} unrolling across the SIM’s layers and iterations, using analytic worst-link SINR gradients, unit-modulus projection, and a small set of learnable step sizes per iteration. The solver preserves hard constraints and monotone descent, offers interpretability, and enables per-frame inference once trained.
\item Extensive simulations on wideband multiuser downlinks demonstrate fast, monotone convergence, a clear layer-depth sweet spot, and consistent gains in worst-link BER and sum rate over digital ZF OFDM-IM and other baselines, which validates the practicality of the proposed model/data-hybrid framework for wideband deployments.
\end{itemize}

The remainder of this paper is organized as follows: Section~\ref{Section2} details the proposed SIM-enhanced multiuser MIMO OFDM-IM system model. The BER-driven optimization problem is then formulated in Section~\ref{Section3}. Section~\ref{Section4} introduces the proposed UPGD-Net and evaluates its complexity. The simulation results and analysis are presented in Section~\ref{Section5}. Finally, conclusions are drawn in Section~\ref{Section6}.

\textbf{Notations:}
In this paper, bold lowercase and uppercase letters are used to denote vectors and matrices, respectively; $ (\mathbf{A})^* $, $ (\mathbf{A})^T $, and $ (\mathbf{A})^H $ represent the conjugate, transpose, and Hermitian transpose of matrix $\mathbf{A}$, respectively; $ |c| $, $ \Re(c) $, and $ \Im(c) $ refer to the magnitude, real part, and imaginary part, respectively, of a complex number $ c $; $ \|\mathbf{A}\|_F $ denotes the Frobenius norm; $ \mathbb{E}(\mathbf{A}) $ stands for the expectation operator; $ \text{diag}(\mathbf{v}) $ produces a diagonal matrix with the elements of vector $ \mathbf{v} $ on the main diagonal; $ \text{vec}(\mathbf{A}) $ denotes the vectorization of a matrix $ \mathbf{A} $; $ \mathbf{A}_{a:b, :} $ and $ \mathbf{A}_{:, c:d} $ represent the submatrices constructed by extracting rows $ a $ to $ b $ and columns $ c $ to $ d $ from matrix $ \mathbf{A} $, respectively; $ \mathbb{C}^{x \times y} $ represents the space of $ x \times y $ complex-valued matrices; $ {\partial f}/{\partial x} $ denotes the partial derivative of a function $ f $ with respect to (w.r.t.) the variable $ x $.

\begin{figure*}
	\centerline{\includegraphics[width=0.95\textwidth]{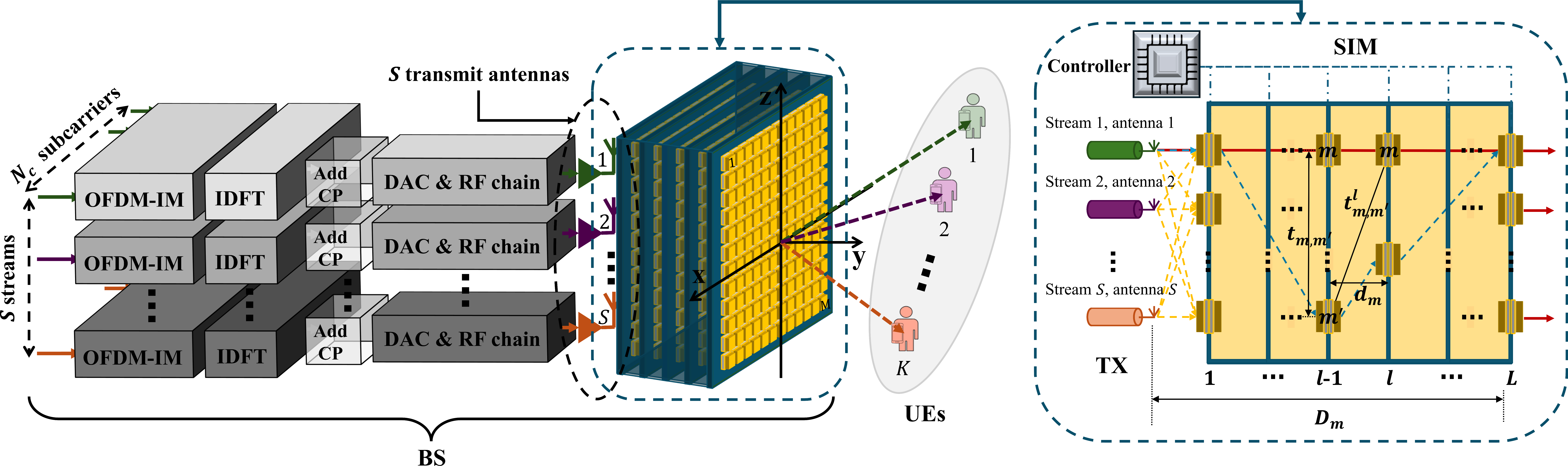}}
	\caption{\centering{Transceiver structure of the SIM-enhanced wideband multiuser MIMO OFDM-IM communication system.}}
	\label{SIM-channel}
\vspace{-0.5 cm}
\end{figure*}

\section{The Proposed SIM-enhanced Wideband Multiuser MIMO OFDM-IM System Model}
\label{Section2}
\subsection{SIM–Enhanced OFDM-IM Transceiver Architecture}
\label{subsec:sim_tx_arch}
~~~As illustrated in Fig.~\ref{SIM-channel}, we focus on the downlink of a wideband multiuser MIMO OFDM system, where the base station (BS) is equipped with a SIM front-end to assist wideband transmission from $S$ transmit antennas to $K$ single-antenna user equipments (UEs) within a specific frequency band. The SIM is designed by an array of $L$ transmissive metasurfaces, each containing $M = M_x \times M_z$ electronically reconfigurable meta-atoms, where $M_x$ and $M_z$ denote the number of meta-atoms along the $x$-axis and $z$-axis, respectively. A smart controller connected to the metasurfaces independently tunes the phase shift of every meta-atom, enabling the SIM to execute wave-domain computing in situ. The large aperture array of the metasurfaces provides the BS additional power margin to surpass $S$ physical antennas at BS. Each stream is processed by one DAC-RF chain and fed to a dedicated transmitter (TX) aperture located immediately behind the first metasurface layer. Hence only $S$ RF chains are required, whose number is far fewer than that of a conventional fully-digital array with $M$ active elements. Unlike conventional hybrid-based wideband OFDM architectures, which require a dedicated precoding vector for each UE, the proposed SIM-enhanced wideband OFDM transceiver architecture employs wave-based computing to enhance signal gain and achieve multiuser fully-analog spatial precoding simultaneously. 

To establish multiple parallel subchannels in the physical space for interference-free transmission, the proposed SIM-enhanced fully-analog system adopts spatial division multiple access (SDMA) with $S\!=\!K$ spatial streams for the sake of simplicity. In contrast to conventional digital SDMA, where each antenna transmits the superposition of digitally precoded symbols, resulting in a high PAPR, the proposed architecture delegates the matrix multiplication to the wave-based computing by SIM. Consequently, each data stream is radiated directly from a dedicated feed antenna, and the BS requires only $N_{TX}=N_{RF}=K$ RF chains. The reduced RF burden, together with the inherently lower PAPR of OFDM-IM, markedly improves power-amplifier efficiency.

Let the frequency point of the $i$-th subcarrier be $f_i = f_0+(i - \frac{N_c+1}{2})\Delta f,~\Delta f = \frac{B_w} {N_c},~i\in \mathcal{N}_c=\{1,\dots,N_c\}$, where $f_0$ is the center frequency, $B_w$ is the bandwidth, and $N_c$ denotes the total number of OFDM subcarriers. For each OFDM symbol, the BS performs the following steps for every stream $k$ belonging to the set of UEs $\mathcal{K}=\{1,\dots,K\}$:
\begin{enumerate}
\item Bit-to-symbol mapping yields a sparse OFDM-IM vector ${\mathbf{x}}_{k}\!=\!\bigl[x_{k}(1),\dots,x_{k}(N_c)\bigr]^{T}\!\in\!\mathbb{C}^{N_{c}}$ containing $N_c$ subcarriers in the $k$-th stream, following the rules in Sec.~\ref{sec:system_IM}.
\item An \(N_{c}\)-point inverse discrete Fourier transform (IDFT) and a cyclic prefix (CP) insertion process convert ${\mathbf{x}}_{k}$ into the time-domain OFDM signal $\tilde{\mathbf{x}}_{k}$.
\item One DAC–RF chain up-converts the waveform and feeds the corresponding antenna, followed by the wave-domain beamforming implemented through the SIM structure.
\item At the receiver (RX), after CP removal and discrete Fourier transform (DFT), each UE performs separate OFDM-IM detection.
\end{enumerate}

Specifically, the spacing between the $m$-th and $m'$-th meta-atoms within the same layer of the SIM is represented as $t_{m, m'}$, while the interlayer spacing between the $m$-th meta-atom on the $l$-th layer and the $m'$-th meta-atom on the ($l\!-\!1$)-th layer is denoted as $t_{m, m'}^l$, where $l\in\mathcal{L}=\{1,\dots,L\}$ and $m,\, m'\in\mathcal{M}=\{1,\dots,M\}$. The axial interlayer distance in the SIM is $d_m$, with the total thickness denoted as $D_m\!=\!d_m\!\times \!L$. The spacing between adjacent meta-atoms within the same layer is denoted as $r_m$. According to the Rayleigh-Sommerfeld theory~\cite{RS}, the wideband transmission coefficients of the multiple layers of SIM on subcarrier $i$ can be expressed as:
\begin{align}
w_{m, m'}^l(i)&=\frac{S_m\,d_m}{{t_{m, m'}^l}^2} (\frac{1}{2\pi\,t_{m, m'}^l}-j\tfrac{f_i}{c})e^{j2\pi t_{m, m'}^lf_i/c},
\label{w1}
\end{align}
where $S_m$ is the area of meta-atom and $c$ is the speed of light. 

Each meta-atom adjusts its transmission coefficient by imposing continuously adjustable phase shifts, represented by $\phi_m^l=e^{j\theta_m^l},~\theta_m^l \in [0,2\pi)$ for the $m$-th meta-atom on the $l$-th layer. The phase-shift matrix of the $l$-th layer is denoted by $\mathbf{\Phi}^l={\rm diag}([\phi_1^l,  \phi_2^l, \ldots, \phi_M^l]^T)\in \mathbb{C}^{M\times M}$. The cumulative effect of signal propagation through multiple layers on subcarrier $i$ is characterized by the transmission functions:
\begin{equation}
\mathbf{G}(i)=\mathbf{\Phi}^L\mathbf{W}^L_i \ldots \mathbf{\Phi}^2\mathbf{W}^2_i\mathbf{\Phi}^1\mathbf{W}^1_i\in \mathbb{C}^{M\times K},
\label{eq:2}
\end{equation}
where $\mathbf{W}^l_i= [w_{m, m'}^l(i)]_{M\times M},~l=2, \ldots, L$ denotes the wideband transmission matrix between the ($l$-$1$)-th layer and the $l$-th layer of SIM and $\mathbf{W}^1_i\in \mathbb{C}^{M\times K}$ describes the transmission from transmit antennas to the first layer of SIM.

{{
\begin{remark}
\textnormal{To establish a fundamental performance benchmark for the proposed SIM-enhanced transceiver, we adopt the assumption of continuous phase shifts with a unit-modulus amplitude response for each meta-atom \cite{PS}. This modeling choice not only facilitates the derivation of a theoretical upper bound but also ensures the differentiability required for optimization. While practical implementation relies on discrete phases, recent studies \cite{SIM-BF4} have validated the inherent robustness of SIM architectures to quantization errors. Specifically, a coarse 2-bit phase resolution incurs only a marginal sum rate penalty ($< 0.5$ bps) compared to the continuous profile, with the performance gap becoming negligible at higher resolutions ($\ge 3$ bits). Consequently, the continuous model serves as a rigorous baseline for the wave-domain beamforming principles explored in this study. Detailed investigations into other hardware non-idealities, such as adjustable magnitudes \cite{PM4} and coupled phase-amplitude tuning mechanisms \cite{PM1}, are reserved for future work.
}
\end{remark}
}}

\begin{remark}
\textnormal{By dynamically adjusting the phase-shift matrix of each metasurface layer, the SIM structure adjusts the EM response in the proposed system to perform two key functions: 
\begin{enumerate}
\item \textbf{Enhancing beamforming gain}: The SIM, equipped at the BS, processes the emitted wavefronts, enhancing the received power at each UE via large-aperture arrays.
\item \textbf{Implementing multiuser precoding}: The SIM spatially separates the signals destined for different UEs, mitigating IUI and optimizing signal reception at each UE.
\end{enumerate}
}
\end{remark}

\subsection{Multiuser OFDM–IM Module Construction at the BS}
\label{sec:system_IM}
\begin{figure}
	\centerline{\includegraphics[width=0.45\textwidth]{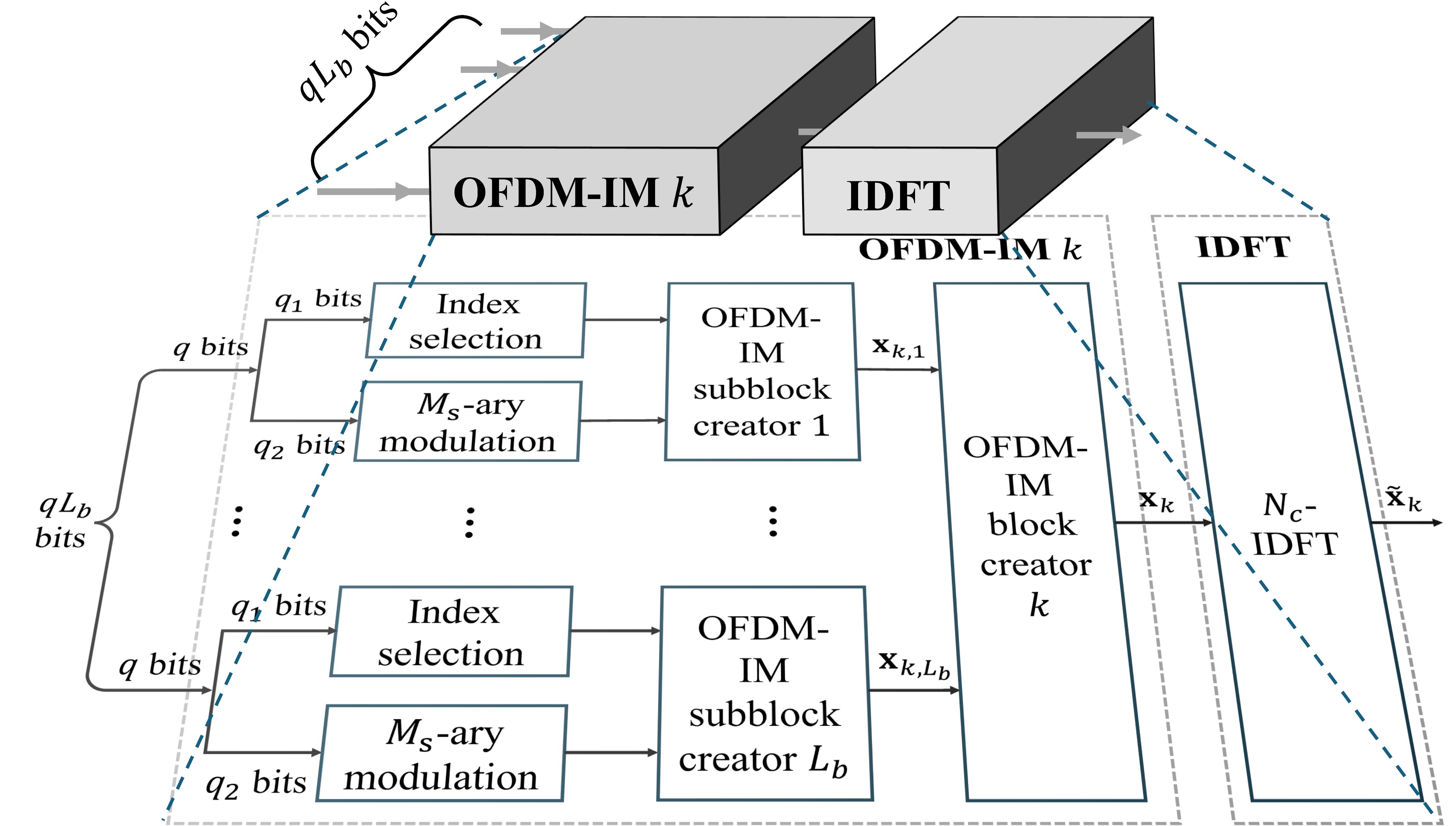}}
	\caption{\centering{OFDM index modulators at the $k$-th branch of the TX.}}
	\label{OFDM-IM}
\vspace{-0.5 cm}
\end{figure}

~~Per OFDM symbol, the available $N_c$ subcarriers are divided into $L_b = N_c/N$ disjoint OFDM-IM subblocks. Each subblock comprises $N$ subcarriers, among which only $V<N$ tones are activated to carry modulated data. The remaining $N-V$ tones remain silent, implicitly conveying index bits. As illustrated in Fig.~\ref{OFDM-IM}, each transmit stream $k\!\in\!\mathcal{K}$ independently maps incoming bitstreams into sparse frequency-domain subblocks via an index selector and a symbol modulator.

For a subblock $\ell\in\mathcal{L}_b=\{1,\dots,L_{b}\}$ of stream $k$, the incoming $q = q_1 + q_2$ bits are split as follows:
\begin{equation}
q_{1}=\Bigl\lfloor\log_{2}\tbinom{N}{V}\Bigr\rfloor\!,
\qquad
q_{2}=V\,\log_{2}M_{s},
\label{eq:bit_split}
\end{equation}
where $q_{1}$ index bits determine the positions of $V$ tones that are active, and $q_{2}$ symbol bits are mapped onto $V$ $M_{s}$–ary modulation data symbols. These $L_b$ subblocks in each transmit stream are aggregated into $q\,L_b$ total bits, which are processed in each branch of the transmitter by the OFDM index modulators.

{{Specifically, we define the selected activation-codebook as
\begin{equation}
\mathcal I_{N,V}
=
\left\{
\mathcal I_1,\mathcal I_2,\ldots,\mathcal I_A
\right\},
\quad
A=2^{q_1},
\end{equation}
where \(\mathcal I_a\subseteq\{1,\ldots,N\}\), \(|\mathcal I_a|=V\), and \(A\leq {N\choose V}\). Each \(q_1\)-bit index word selects one pattern \(\mathcal I_a\) from \(\mathcal I_{N,V}\). }}

Instead of explicitly listing active-tone indices, all activation decisions can be encoded in a global binary activation matrix:
\begin{equation}
\mathbf{Z}\;\triangleq\;\bigl[\,\mathbf{Z}_{1},\dots,\,\mathbf{Z}_{\ell},\dots,\mathbf{Z}_{L_b}\bigr]
\in\{0,1\}^{K\times N_c},
\end{equation}
where $\mathbf{Z}_{\ell} \in \{0,1\}^{K\times N}$ represents the activation decisions in subblock $\ell$. Entry $Z_{k,\ell,n} = 1$ indicates the stream $k$ is active on the $n$-th tone of subblock $\ell$, and $0$ otherwise. To satisfy the designed OFDM-IM structure, each row of $\mathbf{Z}_{\ell}$ must satisfy: $\sum_{n=1}^{N} Z_{k,\ell,n}= V,~\forall k\in \mathcal{K},~\forall \ell \in \mathcal{L}_b$, ensuring that each UE activates exactly $V$ tones per subblock.

Each UE $k$ then maps their $V$ modulated symbols into the selected active tones. The resulting modulated subblock signal is represented compactly as
\begin{equation}
x_{k,\ell,n} = Z_{k,\ell,n} \cdot s_{k,\ell,n},
\label{eq:x_Z_relation}
\end{equation}
where \( s_{k,\ell,n} \in \mathbb{C} \) is nonzero only when \( Z_{k,\ell,n} = 1 \). 

For the $k$-th stream, these sparse subblock vectors $\mathbf{x}_{k,\ell}=\bigl[x_{k,\ell,1},\dots,x_{k,\ell,N}\bigr]^{T}\in \mathbb{C}^{N}$ are then concatenated across all $L_b$ subblocks to produce the full frequency-domain OFDM-IM symbol: $\mathbf{x}_{k}=[\mathbf{x}_{k,1}^T,\dots,\mathbf{x}_{k,L_b}^T]^T\in\mathbb{C}^{N_{c}}$. Finally, an $N_{c}$-point IDFT followed by a CP insertion of $N_{\text{cp}}$ samples transforms $\mathbf{x}_{k}$ into the time-domain signal $\tilde{\mathbf{x}}_{k}$. The resulting signals are simultaneously emitted from the $S$ SIM-equipped antennas after fully-analog precoding.

\begin{remark}
\textnormal{
The use of a unified binary activation matrix $\mathbf{Z}$ offers a compact and flexible representation of OFDM–IM subcarrier activity across users and subblocks. This matrix-centric formulation not only simplifies the modulation and transmission process in classical OFDM-IM systems but also facilitates the joint design of analog precoding, power allocation, and user scheduling. Specifically, $\mathbf{Z}$ serves as the foundational structure for interference-aware SINR computation, fairness-constrained scheduling, and optimization of UE-subcarrier associations in multiuser SIM-enhanced systems.
}
\end{remark}

\subsection{Wideband Channel and Received Signal Model}
~~~We consider a wideband multipath channel model, incorporating multiple scatterers that result in frequency-selective fading effects. {{The tapped delay line channel accounts for $P_k$ resolvable paths between the SIM at the BS and the $k$-th UE. Let $p = 0$ denote the line-of-sight (LoS) path and $p=1,\ldots,P_k$ denote the non-LoS (NLoS) paths. The complex gain and delay associated with the $p$-th path for UE $k$ are denoted as $g_{k,p}$ and $\tau_{k,p}$, respectively.}} The frequency-domain channel vector on subcarrier $i$ is 
\begin{align}
\label{channel}
{{\mathbf{h}_k(i)=  \sum_{p=0}^{P_k}g_{k,p}e^{-j2\pi f_i\tau_{k,p}}{\bm{\alpha}_{k,p}(i)}^H \in \mathbb{C}^{1\times M}, }}
\end{align}
where $\bm{\alpha}_{k,p}(i) \in \mathbb{C}^{M\times1}$ is the steering vector of the uniform planar array corresponding to the elevation angle $\vartheta_{k,p} \in [0,\pi)$ and azimuth angle $\varphi_{k,p}  \in [-\pi/2,\pi/2]$ from the SIM side to UE $k$. The array steering vectors $\bm{\alpha}_{k,p}(i) = \bm{\alpha}_{k,p}^{x}(i) \otimes \bm{\alpha}_{k,p}^{z}(i)$ represent the array response of all meta-atoms on both the $x$-axis and $z$-axis in the $L$-th layer of SIM, which is
\begin{align}
[\bm{\alpha}_{k,p}^{x}(i)] _{m_x}&\triangleq  e^{j2\pi r_{m}\sin(\vartheta_{k,p})\sin(\varphi_{k,p})(m_x-1)f_i/c}, \\
[\bm{\alpha}_{k,p}^{z}(i)] _{m_z}&\triangleq  e^{j2\pi r_{m}\cos(\vartheta_{k,p})(m_z-1)f_i/c},
\end{align}
where $m_x = 1, 2,\ldots, M_x$ and $m_z = 1, 2,\ldots, M_z$. 

Therefore, stacking the wideband channel vectors from BS to all $K$ UEs on subcarrier $i$ yields the global channel matrix:
\begin{align}
\mathbf{H}(i) = \bigl[\mathbf{h}_1(i)^T, \ldots,\mathbf{h}_K(i)^T\bigr]^T \in \mathbb{C}^{K\times M}.
\end{align}

In this paper, we assume perfect channel state information (CSI) at the BS, obtained via a time-division duplexing (TDD) protocol. Following the standard implementation of OFDM-IM, each subblock independently performs index and symbol mapping and is separately detected at the RX using a maximum likelihood (ML) detector. This is justified by the Hermitian Toeplitz structure of the channel correlation matrix, which ensures statistical independence of pairwise error events across subblocks, as established in \cite{IM2}.

{{To relate this global frequency index $i$ to the OFDM–IM structure, there exists a one-to-one mapping between the global subcarrier index $i$ and the local subblock indices $(\ell, n)$:
\begin{equation}
i = (\ell - 1) \cdot N + n,
\quad \text{for } \ell \in \mathcal{L}_b,~n \in \mathcal{N}.
\end{equation}
This index mapping will be used throughout the paper to switch between global frequency index $i$ and subblock notation $(\ell, n)$ for clarity in system modeling.}} Under this setting, the received signal at the $k$-th UE on the $n$-th subcarrier of the $\ell$-th subblock is expressed as:{{
\begin{align}
y_{k,\ell,n}=& \sqrt{p_{k,\ell,n}}\,\mathbf h_{k,\ell,n}\,\mathbf g_{k,\ell,n}\,x_{k,\ell,n}
          \,+\, \nonumber \\ 
&\sum_{j\neq k}\underbrace{\sqrt{p_{j,\ell,n}}\,\mathbf h_{k,\ell,n}\,\mathbf g_{j,\ell,n}}_{\text{Multiuser interference}}\,x_{j,\ell,n}
          +w_{k,\ell,n},
\label{eq:rx-sig}
\end{align}}}
where $p_{k,\ell,n}$ is the power allocated to the $k$-th UE, $\mathbf h_{k,\ell,n} \in \mathbb{C}^{1 \times M}$ is the channel vector from the BS to UE~$k$, and $\mathbf g_{k,\ell,n} \in \mathbb{C}^{M \times 1}$ is the $k$-th column of the analog precoder $\mathbf{G}(i)$ derived from the SIM structure. The noise term $w_{k,\ell,n} \sim \mathcal{C} \mathcal{N}(0,\sigma_w^2)$ represents independent and identically distributed (i.i.d.) additive white Gaussian noise across all UEs, subcarriers, and OFDM-IM subblocks.

Since index bits are embedded implicitly in the nonzero positions and do not receive an extra power budget, all the RF energy is carried only by the $V$ constellation symbols, which are determined by the activation matrix $\mathbf{Z}$. By treating the signals of the remaining UEs as interference according to \eqref{eq:rx-sig}, the SINR $\gamma_{k,\ell,n}$ at the RX of the UE $k$ on the subcarrier $n$ of the subblock $\ell$ is expressed as{{
\begin{equation}
\gamma_{k,\ell, n}=
\frac{Z_{k,\ell,n}\,p_{k,\ell,n}\,|\mathbf h_{k,\ell,n}\,\mathbf g_{k,\ell,n}|^{2}}
     {\displaystyle
      \sum_{j\neq k}Z_{j,\ell,n}\,p_{j,\ell,n}\,
      |\mathbf h_{k,\ell,n}\,\mathbf g_{j,\ell,n}|^{2}
      +\sigma_w^{2}} .
\label{eq:SINR_pwr}
\end{equation}}}

\begin{remark}
\textnormal{
The design of SIM-enhanced OFDM–IM system yields a number of intertwined benefits that are difficult to realize with conventional multiuser MIMO OFDM transceivers.
\begin{enumerate}
\item \textbf{Wave-domain equalization}\,:  
Each meta-atom acts as an adjustable CIR tap; stacking $L$ layers therefore forms a multi-tap analog equalizer. The frequency-selective fading seen by the $N$ tones of every subblock is effectively shaped in the wave domain over the active tones.
\item \textbf{Dual-domain interference suppression}\,:  
The SIM realizes wave-domain ZF-inspired interference suppression and forms $K$-parallel spatial subchannels across all tones, while OFDM-IM supplies frequency-domain index diversity, together suppressing IUI effectively.
\item \textbf{Hardware and energy savings}\,:  
The fully-analog SIM reduces the number of RF chains that burden hybrid or digital architectures, yielding front-end power and cost savings, essential for large-aperture antenna arrays.
\item \textbf{Scalable optimization}\,:  
Encoding the subcarrier map in the activation matrix $\mathbf Z$ confines the phase-shift optimization to a low-dimensional wave-domain optimization problem, enabling scalable design over ultra-wide bandwidths.
\item \textbf{Lower PAPR and PA relief}\,:  
Wave-domain ZF precoding transmits each stream via a single antenna, inherently reducing the PAPR. The $N\!-\!V$ deliberately silent sub-tones per OFDM-IM subblock further reduce the probability of instantaneous peaks.
\end{enumerate}
}
\end{remark}

{{
\begin{remark}
\textnormal{The validity of the forward-propagation model is verified by the specific hardware architecture adopted in this work, consistent with the transmissive SIM prototypes reported in~\cite{SIM1}. By utilizing meta-atoms optimized for unidirectional transmission, the system inherently suppresses reverse coupling between layers and self-oscillation, ensuring that EM energy propagates primarily in the forward direction. Structurally, the metasurface stack is enclosed within a shielded metallic chassis lined with absorptive materials to mitigate edge/side diffraction and lateral leakage. Furthermore, the cumulative energy of higher-order reflections between layers attenuates exponentially, and an additional anti-reflection coating is integrated at the back of the stack to eliminate residual backward scattering. As such, we do not consider backward and anomalous waves in our theoretical model. Nonetheless, future designs could intentionally leverage controlled inter-layer feedback to realize wave-domain recursive networks.
}
\end{remark}
}}

{{ 
\begin{remark}
\textnormal{We acknowledge that the analytical transmission models in \eqref{w1}, \eqref{eq:2} and \eqref{channel} may exhibit discrepancies when compared to practical hardware implementations. These deviations arise not only from inevitable fabrication tolerances but also from simplified physical assumptions, specifically the omission of resistive/dielectric losses and intra-layer mutual coupling effects. To bridge this gap, an in-situ calibration procedure is essential before the SIM’s practical deployment. By transmitting pilot sequences and measuring the received response for the end-to-end learning, the effective transmission coefficients can be iteratively refined using error back-propagation algorithms~\cite{DL}. This data-driven update naturally compensates for the aforementioned modeling mismatches and hardware imperfections. Given that this work focuses on the fundamental precoding framework in MIMO OFDM-IM systems, the detailed implementation of such calibration schemes is designated for future work.
}
\end{remark}
}}

\section{Optimization Problem Formulation}
\label{Section3}
~~~In this section, we investigate the design of SIM phase shifts to minimize the BER of the proposed OFDM-IM system. Unlike conventional OFDM systems where all subcarriers are active and metasurface coefficient optimization typically targets sum rate maximization, we consider a more practical BER metric for system design. To tackle the analytical complexity of BER expressions in SIM-enhanced OFDM-IM, we next reformulate the BER optimization problem into a max-min optimization problem.

\subsection{Single-Subblock BER Derivation}
~~~Let $\mathbf x_{k,\ell}$ denote the transmitted vector in the frequency domain and $\widehat{\mathbf x}_{k,\ell}$ denote the signal detected erroneously by the RX in valid indices and constellation symbols. Here $x_{k,\ell,n},\widehat x_{k,\ell,n}\in \mathcal{S}\cup\{0\}$, where \(\mathcal{S}\) denotes the adopted \(M_s\)-ary constellation alphabet and \(0\) models an inactive tone in OFDM-IM. For a fixed channel under the TDD protocol, the conditional pairwise error probability (CPEP) in \cite{CPEP} collapses to the deterministic pairwise error probability (PEP) form between two OFDM-IM blocks $\mathbf x_{k,\ell}$ and $\widehat{\mathbf x}_{k,\ell}$ of UE~$k$ in the $\ell$-th subblock

\begin{equation}
P\!\bigl(\mathbf x_{k,\ell}\!\rightarrow\!\widehat{\mathbf x}_{k,\ell}\bigr)=
Q\!\Bigl(
  \sqrt{\tfrac12
        \sum_{n=1}^N
        \gamma_{k,\ell,n}\,
        \bigl|x_{k,\ell,n}-\widehat x_{k,\ell,n}\bigr|^{2}}
     \Bigr),
\label{eq:PEP-k}
\end{equation}
where $Q$-function is $Q(x)=\tfrac1\pi\!\int_{0}^{\pi/2}\!\exp\!(-x^{2}/(2\sin^{2}\zeta))d\zeta$.

{{Let $\mathcal X_{k,\ell}$ denote the selected OFDM-IM codebook for UE $k$ in subblock $\ell$:
\begin{equation}
\mathcal X_{k,\ell}
=
\left\{
\mathbf x(\mathcal I_a,\mathbf s)
\mid
\mathcal I_a\in\mathcal I_{N,V},
\mathbf s\in\mathcal S^{V}
\right\},
\end{equation}
and its cardinality is $n_{k,\ell}=|\mathcal X_{k,\ell}|=2^{q_1}M_s^V$ \cite{IM1}.}} According to \cite{IM2}, the union upper bound of BER can be expressed as
\begin{equation}
P_{b,k,\ell}=
\frac{1}{q\,n_{k,\ell}}
\sum_{\mathbf x_{k,\ell}}\sum_{\widehat{\mathbf x}_{k,\ell}}
e\bigl(\mathbf x_{k,\ell},\widehat{\mathbf x}_{k,\ell}\bigr)\;
P\bigl(\mathbf x_{k,\ell}\!\rightarrow\!\widehat{\mathbf x}_{k,\ell}\bigr),
\label{eq:ABEP-union}
\end{equation}
where $e\bigl(\mathbf x_{k,\ell},\widehat{\mathbf x}_{k,\ell}\bigr)$ denotes the Hamming weight between the two bit labels for the corresponding pairwise error event.

\begin{table*}[h]
\centering
\caption{\centering{\protect\\{\textsc{Three representative dominant error classes used in the BER surrogate derivation.}}}}
\setlength{\tabcolsep}{1.2pt}
\renewcommand\arraystretch{2}
\label{tab:OFDMIM_error_classes}
\begin{tabular}{|m{1cm}<{\centering}|m{7cm}|m{1.5cm}<{\centering}|m{1.5cm}<{\centering}|m{1.6cm}<{\centering}|m{5.2cm}<{\centering}|}
\hline
\textbf{Class} &
\textbf{~~~~~~~~~~~~~~~~~~~~~~~~Error pattern} &
\makecell{\textbf{Subcarriers}\\ \textbf{affected}} &
\makecell{\textbf{Squared} \\ \textbf{distance} \\ $d_c^2$} &
\makecell{\textbf{Pair} \\ \textbf{multiplicity}\\ $N_{k,\ell,c}$} &
\makecell{\textbf{Bit error weight}\\ $w_{k,\ell,c}$} \\ \hline

$\textbf{C}_1$ &
One active index is identified as inactive; the constellation symbol is unchanged. &
1 &
$E_s$ &
\multirow{3}{=}{\centering \\{{$\left|\mathcal{C}_{k,\ell,c}\right|$}}} &
\multirow{3}{=}{\centering \\{{$\frac{1}{N_{k,\ell,c}}
\sum_{(\mathbf x_{k,\ell},\,\widehat{\mathbf x}_{k,\ell})\in \mathcal C_{k,\ell,c}}
e\bigl(\mathbf x_{k,\ell},\widehat{\mathbf x}_{k,\ell}\bigr)$}}} \\ \cline{1-4}

$\textbf{C}_2$ &
Constellation symbol on an active subcarrier is identified as one of its nearest neighbors; the index set is unchanged. &
1 &
$d_{\min}^{(\mathcal S)}$ &
&
\\ \cline{1-4}

$\textbf{C}_3$ &
Within the same subcarrier: index error and the recognized symbol is the nearest neighbor of the transmitted symbol. &
1 &
$E_s+d_{\min}^{(\mathcal S)}$ &
&
\\ \hline

\multicolumn{6}{|p{18cm}|}{
~~\textbf{Note:}
$E_s=\mathbb E[|s|^{2}]$ is the average symbol energy;
$d_{\min}^{(\mathcal S)}$ is the minimum squared Euclidean distance between constellation nearest neighbors.
{{Since the practical OFDM-IM mapper employs only a lookup-table subset of size $2^{q_1}$, the pair multiplicities $N_{k,\ell,c}$ and the bit-error weights $w_{k,\ell,c}$ are computed offline from the adopted lookup-table codebook rather than from the full combinatorial activation set.}}} \\ \hline
\end{tabular}
\vspace{-0.5 cm}
\end{table*}

Since the BER in~\eqref{eq:ABEP-union} for OFDM-IM is hard to obtain, we construct a dominant-error approximation by keeping the three classes with minimum distance. The others have larger Euclidean distances, and their PEP terms decay exponentially with SINR and can be neglected in the analytical upper bound. {{For each dominant error class $c\in\mathcal{C}=\{C_1,C_2,C_3\}$, define the corresponding ordered codeword-pair set
\begin{equation}
\mathcal{C}_{k,\ell,c}
\subseteq
\mathcal{X}_{k,\ell} \times \mathcal{X}_{k,\ell},
\qquad
\mathbf x_{k,\ell}\neq \widehat{\mathbf x}_{k,\ell},
\end{equation}
where each pair $(\mathbf x_{k,\ell},\widehat{\mathbf x}_{k,\ell})\in\mathcal{C}_{k,\ell,c}$ belongs to the dominant error class $c$ listed in TABLE~\ref{tab:OFDMIM_error_classes}. Here, $\beta_c=d_c^2$ is the class-dependent squared Euclidean distance. The pair multiplicity $N_{k,\ell,c}$ of class $c$ is defined as the number of ordered codeword pairs in the adopted lookup-table codebook. The average bit error weight $w_{k,\ell,c}$ of class~$c$ is defined as the average Hamming distance over all ordered codeword pairs in $\mathcal C_{k,\ell,c}$. }}

Since all pairs in $\mathcal C_{k,\ell,c}$ belong to the same dominant squared distance class, their aggregate contribution in \eqref{eq:ABEP-union} can be approximated by a representative class-level PEP:
\begin{align}
P_{b,k,\ell}=&\sum_{c\in\mathcal C}
\frac{1}{q\,n_{k,\ell}}  \nonumber \\ 
&\sum_{(\mathbf x_{k,\ell},\,\widehat{\mathbf x}_{k,\ell})\in \mathcal C_{k,\ell,c}}
e\bigl(\mathbf x_{k,\ell},\widehat{\mathbf x}_{k,\ell}\bigr)\;
P\bigl(\mathbf x_{k,\ell}\!\rightarrow\!\widehat{\mathbf x}_{k,\ell}\bigr), \\
&=\sum_{c\in\mathcal C}
\frac{1}{q\,n_{k,\ell}} N_{k,\ell,c}\,w_{k,\ell,c}\, P_{\text{PEP},c},
\label{eq:ABEP-union-k}
\end{align}
where $P_{\text{PEP},c}$ denotes the representative class-level PEP. Using Craig's representation, $P_{\text{PEP},c}$ is given by
\begin{align}
P_{\text{PEP},c}
  =\frac{1}{\pi}\int_{0}^{\pi/2}\!\!
    \exp\!\Bigl(
       -\frac{\beta_{c}}{4\sin^{2}\zeta}
        \sum_{n=1}^N\gamma_{k,\ell,n}\, \delta_{k,\ell,n, c}^{2}
    \Bigr)d\zeta,
\label{eq:PEP_class}
\end{align}
where $\delta_{k,\ell,n,c}^{2}$ denotes the normalized representative per-subcarrier distance profile of class $c$, with $\sum_{n=1}^{N}\delta_{k,\ell,n,c}^{2}=1$.

Both quantities $N_{k,\ell,c}$ and $w_{k,\ell,c}$ depend only on the codebook structure and the specific bit mapping used for index and symbol modulation. They are independent of all optimization variables, including precoder $\mathbf G$,~power allocation~$\mathbf p$, and index–activation matrix $\mathbf Z$. Hence, \eqref{eq:ABEP-union-k} can be rewritten as
\begin{align}
P_{b,k,\ell}\,=&\int_{0}^{\pi/2}\,
         \sum_{c\in\mathcal C}\,
         \underbrace{\frac{N_{k,\ell,c}\,w_{k,\ell,c}}{\pi\,q \,n_{k,\ell}}}_{W_{k,\ell,c}>0}  \nonumber \\ 
& \exp\Bigl(
  -\frac{\beta_c}{4\sin^{2}\zeta}
   \sum_{n=1}^N\gamma_{k,\ell,n}\,
        \delta_{k,\ell,n,c}^{2}
  \Bigr)d\zeta , \label{BER}
\end{align}
where $W_{k,\ell,c}$ is a positive scaling constant determined by the codebook structure and the chosen Gray mapping. Since the objective \eqref{BER} is a non-increasing function w.r.t. $\gamma_{k,\ell,n}$, these constants do not affect the direction of optimization.

\subsection{Network-Level Worst‑Link BER and SINR}
~~Let $P_{b,k}$ denote the BER of UE $k$ over one OFDM frame. Since the $L_b$ subblocks share the same OFDM-IM parameters, the BER derivation can be carried out on a representative subblock in the statistical-average sense. Therefore,
\begin{equation}
P_{b,k}=\frac{1}{L_b}\sum_{\ell=1}^{L_b}P_{b,k,\ell},
\quad
\mathbb{E}[P_{b,k,\ell}] = \mathbb{E}[P_{b,k,\ell'}],\quad \forall \ell,\ell'.
\end{equation}

Accordingly, the following surrogate derivation focuses on a generic subblock without implying sample-wise equality across all subblocks. Using the representative-subblock approximation in the statistical-average sense, we approximate the worst-user BER objective as
\begin{align}
P_b^{\max} &\,=\, \max_{k} P_{b,k,\ell} \nonumber \\ 
& \,=\, \max_{k} \int_{0}^{\pi/2}\,
         \sum_{c\in\mathcal C}\,W_{k,\ell,c}
\exp\Bigl(
  -\frac{\beta_c}{4\sin^{2}\zeta}
   \Sigma_{k,\ell,c}
  \Bigr)
d\zeta ,
\label{eq:abep_max}
\end{align}
where $\Sigma_{k,\ell,c}=\sum_{n=1}^N\gamma_{k,\ell,n}\,\delta_{k,\ell,n,c}^{2}$. Because the overall objective is a positive linear combination of monotonically decreasing exponentials, these positive constants $W_{k,\ell,c}$ do not affect the direction of optimization in \eqref{eq:abep_max}.

Due to the normalized distance-profile definition in (20), 
\(\Upsilon_{k,\ell,c}=1\). We keep \(\Upsilon_{k,\ell,c}\) in the following
derivation to explicitly show the role of the class-level distance profile.

Let the error exponent be $E_{k,\ell,c}\;\triangleq\;\exp\!\Bigl(  -\tfrac{\beta_{c}}{4\sin^{2}\zeta}\Sigma_{k,\ell,c}\Bigr)$, $  \gamma_{\min}^{(c)}\triangleq\min_{n:\delta^2_{k,\ell,n,c}>0}\,\gamma_{k,\ell,n}$, and $  \Upsilon_{k,\ell,c} \triangleq\sum_{n=1}^N\delta_{k,\ell,n,c}^{2}=1$. For every UE $k$ and class $c$, $\Sigma_{k,\ell,c}\ge  \gamma^{(c)}_{\min}\,\Upsilon_{k,\ell,c}
$. Hence,
\begin{equation}
E_{k,\ell,c}\;\le\;
\exp\!\Bigl(
 -\tfrac{\beta_{c}}{4\sin^{2}\zeta}\,
  \gamma_{\min}^{(c)}\,\Upsilon_{k,\ell,c}
\Bigr).
\label{eq:exp_step1}
\end{equation}

Define the minimum distance parameter \( \beta_{\min}=\min_{c\in\mathcal C}\beta_{c} \) and
\( \eta_{c}=\gamma_{\min}^{(c)}/\beta_{\min} \). By scaling the effective signal power term $\beta_{c}\,\gamma_{\min}^{(c)}$ using the lower bound $\beta_{c} \ge \beta_{\min}$, we obtain:
\begin{align}
\beta_{c}\,\gamma_{\min}^{(c)}
= \beta_{c}\,\frac{\beta_{\min}}{\beta_{\min}}\,\gamma_{\min}^{(c)}
\ \ge\
\beta_{\min}^{2}\,\frac{\gamma_{\min}^{(c)}}{\beta_{\min}}.
\label{eq:exp_step2}
\end{align}

Taking the minimum over the three classes \(c\in\mathcal C\) gives
\begin{align}
\eta
=\min_{c\in\mathcal C}\eta_{c}
=\min_{c\in\mathcal C}\frac{\gamma_{\min}^{(c)}}{\beta_{\min}}
=\frac{1}{\beta_{\min}}\min_{c\in\mathcal C}\min_{n:\delta^2_{k,\ell,n,c}>0}\gamma_{k,\ell,n}.
\label{eq:exp_step3}
\end{align}

Each class \(c\) contributes an exponent \(E_{k,\ell,c}\) to the union bound. 
Lower-bounding \( \Sigma_{k,\ell,c} \) by its worst–link term and using 
\( \beta_{c}\ge\beta_{\min} \) (replacing \( \beta_c \) by \( \beta_{\min} \)) yield
\begin{equation}
E_{k,\ell,c}\ \le\
\exp\!\Bigl(
 -\tfrac{\beta_{\min}^2\,\eta_{c}}{4\sin^{2}\zeta}\,\Upsilon_{k,\ell,c}
\Bigr)\ \le\
\exp\!\Bigl(
 -\tfrac{\beta_{\min}^2\,\eta}{4\sin^{2}\zeta}\,\Upsilon_{k,\ell,c}
\Bigr).
\label{eq:exp_final}
\end{equation}
Since each \(E_{k,\ell,c}\) is a decreasing function of \(\eta\), and summation over \(c\) and integration over \(\zeta\) preserve monotonicity (all weights are positive), the worst-link union-bound BER is also monotonically decreasing with respect to (w.r.t.) \(\eta\).

{{Define the active-link set induced by the OFDM-IM activation matrix as $\mathcal{A}(\mathbf{Z}) \triangleq \{(k,\ell,n)\mid Z_{k,\ell,n}=1\}$. For the three dominant error classes, the representative distance profile is supported by the active tones involved in OFDM-IM symbol or index decisions. Optimizing the minimum SINR over all active links provides a conservative common surrogate for all class-specific worst-tone SINRs.}} Therefore, the original BER minimization problem is cast into the following surrogate optimization problem:
\begin{align}
  \min_{\{\theta_m^l\}}\;P_{b}^{\max}
  \xrightarrow{\text{Surrogate}}
  \max_{\{\theta_m^l\}}
       \eta= \frac1{\beta_{\min}}\min_{(k,\ell,n)\in\mathcal{A}(\mathbf{Z})}\gamma_{k,\ell,n}.
\end{align}

\begin{remark}
\textnormal{Three dominant error classes in TABLE~\ref{tab:OFDMIM_error_classes} have different constant Euclidean-distance factors, which are determined only by the adopted OFDM-IM lookup table and modulation alphabet. These parameters are independent of the channel realization, precoder design, or power allocation. These constants affect the tightness of the analytical BER approximation but do not alter the monotonic relationship between the BER bound and the active-link SINRs. Therefore, we use the worst active-link SINR as a conservative and tractable surrogate for BER-driven SIM phase optimization.}
\end{remark}

\subsection{Final Optimization Problem}
~~~Here we start with a BER optimization problem based on the proposed  framework under the representative-subblock notation described above, subject to (s.t.) the non-convex unit-modulus constraints:
\begin{subequations}
\label{eq:opt_frame}
\begin{align}
\label{Opt}
\mathcal{P}_1:~\min_{\{\theta_m^l\}}~&P_{b}^{\max} \\
s.t.\quad 
     & \mathbf{G}_n=\mathbf{\Phi}^L\mathbf{W}^L_n \ldots \mathbf{\Phi}^2\mathbf{W}^2_n\mathbf{\Phi}^1\mathbf{W}^1_n, \label{st.Pi}\\
     &\mathbf{\Phi}^l={\rm diag}([\phi_1^l,  \phi_2^l, \ldots, \phi_M^l]^T),~\forall l\in \mathcal{L}, \label{st.pPhi} \\
     &|\phi_m^l|=|e^{j\theta_m^l}|=1,~m\in \mathcal{M},~\forall l\in \mathcal{L},  \label{st.phi}
\end{align}
\end{subequations}

{{Collecting the SINR constraints for all active links $(k,\ell,n)\in\mathcal{A}(\mathbf{Z})$, we simplify the objective function from Problem $\mathcal{P}_1$ to Problem $\mathcal{P}_2$:
\begin{subequations}
\label{eq:opt_frame}
\begin{align}
\label{Opt}
\mathcal{P}_2:~~\max_{\{\theta_m^l\}}~&\min_{(k,\ell,n)\in\mathcal{A}(\mathbf{Z})}\,
\gamma_{k,\ell, n} \\
s.t.\quad 
     & \eqref{st.Pi},~\eqref{st.pPhi},~\eqref{st.phi}.
\end{align}
\end{subequations}
where the optimization problem $\mathcal{P}_2$ aims to maximize the worst-link SINR across all UEs and active subcarriers by tuning the phase shifts of a fully-analog SIM-enhanced precoder. $\mathbf{Z}$ is a fixed input parameter determined by the data source, $\mathbf p$ is a precomputed feasible power allocation satisfying the total power budget \(P_t\), and the optimization is performed over ${\theta_m^l}$.}}

However, it is challenging to solve the max-min problem due to the nonlinear and highly coupled structure of the SINR expression w.r.t. ${\theta_m^l}$. Furthermore, in OFDM-IM, each transmitted bitstream activates a distinct subset of subcarriers according to the index bits, which corresponds to a unique subcarrier activation matrix $\mathbf{Z}$. In traditional solvers, every change in the subcarrier activation matrix $\mathbf{Z}$ requires solving the full max-min SINR optimization problem in \eqref{eq:opt_frame} from scratch. This is computationally expensive and impractical for real-time OFDM-IM transmission where $\mathbf{Z}$ may be frequently re-optimized due to UE scheduling or adaptive resource allocation. To address this challenge, we propose a DUN that embeds the structure of the PGD method into a trainable deep neural architecture in Section~\ref{Section4} to obtain the SIM phase solution. 

\section{DUN Solution for SIM Phase Optimization}
\label{Section4}
~~~In this section, we propose a deep UPGD-Net to efficiently approximate the optimal phase configuration under varying subcarrier activation patterns $\mathbf{Z}$ in our proposed SIM-enhanced multiuser OFDM-IM system. The proposed UPGD-Net acquires a data-driven mapping from the subcarrier assignment to the near-optimal SIM configuration.

\subsection{Motivation and Design Principle}
~~~The max–min SINR optimization problem in \eqref{eq:opt_frame} involves a cascade of $L\times M$ unit--modulus phase matrices and must be resolved on a subcarrier basis each time the OFDM-IM activation matrix $\mathbf Z$ changes. Running conventional PGD online is therefore impractical: The iteration has to start from scratch for every new $ (\mathbf H,\mathbf Z)$ pair, and its convergence speed is governed by hand–tuned step sizes.

DUN offers a principled remedy. A finite PGD trajectory can be rewritten as a feed-forward composition of deterministic operators, where the projection, the SINR gradient, and the SIM matrix product are all differentiable. The proposed UPGD-Net is trained offline by back-propagating through the $T$ unrolled stages so that the learned step sizes minimize the statistical loss $-\mathbb E[\gamma_{\min}]$ on a collection of channels and activation patterns. Once trained, UPGD-Net executes exactly $T$ matrix–vector operations and analytical gradient evaluations, which inherits the physical interpretability of PGD but becomes amenable to data-driven acceleration.

The proposed architecture is fundamentally different from a closed box DNN. The proposed SIM-enhanced OFDM-IM system features: (i) SIM's multilayer metasurface cascades, (ii) per-subcarrier SINR coupling, and (iii) a max-min objective that depends on the instantaneous $\mathbf Z$. Each of these traits is naturally captured by UPGD-Net:
\begin{enumerate}
{{\item Coupled physical-algorithmic unrolling: The nested computational graph aligns the algorithmic iterations~$t$ with the physical wave propagation through layer indices~$l$, enforcing physics-compliant gradient back-propagation;}}
\item The worst-link SINR gradient, shared across layers, encapsulates the cross-subcarrier interference structure;
\item Trainable step sizes shorten the convergence path, yielding real-time fully-analog beamforming in dynamic OFDM-IM links.
\end{enumerate}

In summary, UPGD-Net preserves the convergence behavior of the analytical solver, respects the hardware constraints of the multilayer SIM, and generalizes to any activation pattern without online re-optimization thanks to training on a diverse set of $\mathbf Z$ matrices. It bridges the gap between model-based optimization and data-driven acceleration~\cite{DUN1}, providing an interpretable, scalable, and latency-free solution to the SIM phase problem formulated in Section~\ref{Section3}. 

\subsection{UPGD Network Architecture}
\begin{figure*}
	\centerline{\includegraphics[width=0.9\textwidth]{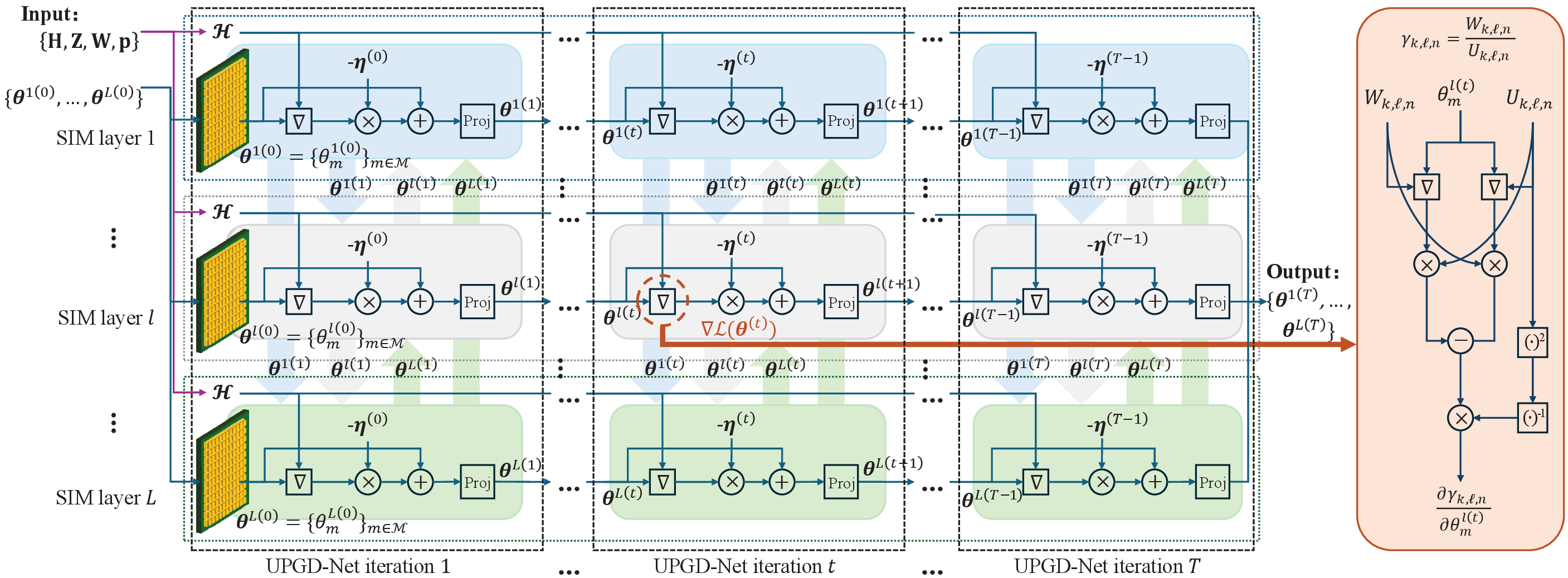}}
\vspace{0.2 cm}
	\caption{{UPGD-Net for optimizing the phase shifts of the $L$-layer SIM over $T$ projected-gradient iterations. Columns denote successive iterations $t=0:T-1$, and rows correspond to SIM layers $l=1:L$. In each cell, the analytic gradient $\nabla$ is scaled by the trainable step size $-\eta^{(t)}$, added to the current phase-shift tensors, and then projected to $[0,2\pi)$. A vertical coupling bus delivers the worst-link SINR gradient, computed from the cascade precoder of all layers, to every $\nabla$-block within the same iteration. All modules are model-driven and fully differentiable.}}
\label{UPGD}
\vspace{-0.5 cm}
\end{figure*}

~~~As shown in Fig. \ref{UPGD}, the proposed UPGD-Net is organized as a two-dimensional grid: the horizontal axis unrolls the $T$ projected–gradient iterations ($t=0,\ldots ,T\!-\!1$), whereas the vertical axis stacks the $L$ physical SIM layers ($l=1,\ldots ,L$). A single grid cell therefore captures the micro-update for the layer~$l$ executed at iteration~$t$. The forward and backward flows are described as follows.

\subsubsection{Forward Propagation:}
The inputs of the proposed UPGD-Net are: (i) the input tuple $\mathcal{H}=\{\mathbf H, \mathbf Z, \mathbf W, \mathbf p\}$ for SINR calculation, which contains the channel matrix $\mathbf H$, activation matrix $\mathbf Z$, SIM transmission matrix $\mathbf W =\{W_n^{(l)}\}$, and the power allocation matrix $\mathbf p$;  (ii) the initialized phase-shift tensor ${\boldsymbol{\Theta}}^{(0)}=\{{\boldsymbol{\theta}}^{1(0)}, \ldots, {\boldsymbol{\theta}}^{L(0)}\}$, where $ {\boldsymbol{\theta}}^{l(0)}={\{\theta_m^{l(0)}\}}_{m\in \mathcal{M}}$. In each iteration $t$, the following computations are performed:

\begin{enumerate}[Step 1]
\item\textbf{SIM precoder construction:}
For the $n$-th local subcarrier in the $\ell$-th OFDM-IM subblock, the phase-shift tensor $\{\boldsymbol{\theta}^{l(t)}\}_{l\in \mathcal{L}}$ at every iteration $t$ is used to form the cascade fully-analog precoder. Since all OFDM-IM subblocks are statistically independent and identically parameterized, with a slight abuse of notation, we suppress the subblock index $\ell$ and write $\mathbf G_n^{(t)}=\prod_{l=L}^{1}\mathbf\Phi^{l(t)}\mathbf W_n^l$, where $n$ denotes the local subcarrier index within a generic subblock.
    \item \textbf{SINR computation:} Using $\mathbf G_n^{(t)}$, $\mathbf H$, $\mathbf Z$, and $\mathbf p$, the SINRs $\gamma_{k,\ell,n}^{(t)}$ for different subcarriers are evaluated in parallel. The instantaneous SINR for each UE-subcarrier pair $(k, \ell, n)$ is evaluated as \eqref{eq:SINR_pwr}.
    \item \textbf{Loss function:} Maximizing the minimum is equivalent to minimizing its negative, so the worst-link SINR is then used to compute the sub-gradient of the loss function:{{
\begin{equation}
\mathcal{L}^{(t)} = -\min_{(k,\ell,n)\in\mathcal{A}(\mathbf{Z})} \gamma_{k,\ell, n}^{(t)},
\label{loss-fun}
\end{equation}}}
where $(k^\ast,\ell^\ast,n^\ast)=\arg\min_{(k,\ell,n)\in\mathcal{A}(\mathbf{Z})}\gamma_{k,\ell,n}^{(t)}$ denotes the worst-case UE-subcarrier pair.
    \item \textbf{Gradient evaluation:}
    The partial derivative of the SINR w.r.t. each phase shift $\theta_m^{l(t)}$ at the $m$-th meta-atom on the $l$-th layer is given by \emph{Lemma}~\ref{lemma1}.
\begin{lemma}
\label{lemma1}{{
\textnormal{Let $a_{k,\ell,n}=\mathbf h_{k,\ell,n}\mathbf g_{k,\ell,n}$ and $b_{k,j,\ell,n}=\mathbf h_{k,\ell,n}\mathbf g_{j,\ell,n}$. Define $W_{k,\ell,n}=Z_{k,\ell,n}p_{k,\ell,n}|a_{k,\ell,n}|^2$ and $U_{k,\ell,n}=\sum_{j\ne k}Z_{j,\ell,n}p_{j,\ell,n}|b_{k,j,\ell,n}|^2+\sigma_w^2$. For the $m$-th meta-atom's phase of layer $l$ at iteration $t$, the analytic form of the gradient is derived via chain rule:
  \begin{equation}
  \frac{\partial \gamma_{k,\ell,n}^{(t)}}{\partial \theta_m^{l(t)}} = \frac{1}{U_{k,\ell,n}^2} \left( U_{k,\ell,n} \frac{\partial W_{k,\ell,n}}{\partial \theta_m^{l(t)}} - W_{k,\ell,n} \frac{\partial U_{k,\ell,n}}{\partial \theta_m^{l(t)}} \right).
\label{Grad}
  \end{equation}}}
}
\end{lemma}
\begin{proof}
\textnormal{Please see Appendix A.}
\renewcommand{\qedsymbol}{}
\end{proof}
    \item \textbf{PGD update and projection:} The SIM phase-shift tensors are updated using a trainable step size $\eta^{(t)}$, followed by projection:
    \begin{equation}
        \theta_m^{l(t+1)} = {\text{ Proj}}_{[0, 2\pi)}\left[ \theta_m^{l(t)} - \eta^{(t)} \cdot \frac{\partial \mathcal{L}^{(t)}}{\partial \theta_m^{l(t)}} \right].
    \end{equation}
The projection enforces the unit-modulus constraint and the updated phases enter the next iteration~$t{+}1$. After $T$ stages, the network outputs $\theta_m^{l(T)}$, which is directly applied to the SIM controller; no run-time looping is required. After updating $L\times M$ meta-atom phase shifts, the network outputs the optimized SIM phase configuration $\boldsymbol{\Theta}^{(T)}$ for final deployment.
\end{enumerate}

\subsubsection{Backward Propagation:}

The network parameters to be trained are the per-iteration, layer-shared step sizes $\{\eta^{(t)}\}_{t=0}^{T-1}$. During training, the overall loss is computed as:
\begin{equation}
    \mathcal{L}_{\mathrm{total}} = \mathcal{L}^{(T)} = - \min_{(k,\ell,n)\in\mathcal{A}(\mathbf{Z})}\, \gamma_{k,\ell,n}^{(T)}.
\label{lf}
\end{equation}

Using backpropagation, the gradients of $\mathcal{L}_{\mathrm{total}}$ are propagated backward through the $T$ unfolded layers via the chain rule. Each trainable step size $\eta^{(t)}$ is updated by:
\begin{equation}
    \frac{\partial \mathcal{L}_{\mathrm{total}}}{\partial \eta^{(t)}} = \sum_{m, l} \frac{\partial \mathcal{L}_{\mathrm{total}}}{\partial \theta_m^{l(t+1)}} \cdot \left( -\frac{\partial \mathcal{L}^{(t)}}{\partial \theta_m^{l(t)}} \right).
\label{backpropa}
\end{equation}

The Adam optimizer is exploited to update the step sizes, while all model-driven blocks remain weight-free and fully differentiable. All updates are efficiently implemented using modern automatic differentiation frameworks, leveraging the fully differentiable structure of UPGD-Net.

{{
The end-to-end training with per-iteration shared steps is demonstrated in Algorithm~\ref{ag:1}. Each forward pass unrolls $T$ projected-gradient steps, and the gradient is recomputed in the next iteration using the previous one. After $T$ steps, the final test loss {{$-\min_{(k,\ell,n)\in\mathcal A(\mathbf Z)}\gamma_{k,\ell,n}^{(T)}$}} is backpropagated through the unrolled structure to update step sizes, while phase shifts are re-optimized per frame by the unrolled solver and are not learned across data. At inference time the epoch loop and backprop are disabled, we freeze $\{\eta^{(t)}\}_{t=0}^{T-1}$ and run the forward unroll once, which yields a near-optimal phase shift tensor in $T$ lightweight updates. 
}}

\begin{algorithm}
\caption{UPGD-Net: End-to-end training with per-iteration shared steps}
\label{ag:1}
\textbf{Initialization:} Phase shifts $\boldsymbol{\Theta}^{(0)}$, channels $\mathbf H$, per-layer transmissions $\mathbf W$,
activation matrix $\mathbf Z$, power allocation matrix $\mathbf p$, layers $L$, training epoch $E$, iterations $T$, trainable step sizes $\{\eta^{(t)}\}_{t=0}^{T-1}$. \\
\begin{algorithmic}[1]
\FOR{$ epoch =1$ \TO $E$}
\FOR{$t=0$ \TO $T-1$}
\STATE \textit{// Form cascade with current phases $\boldsymbol{\Theta}^{(t)}$}
  \STATE For each subcarrier $n$, form $\mathbf G_n^{(t)}=\prod_{l=L}^{1}\mathbf \Phi^{l(t)} \mathbf W_n^{l}$, with $\mathbf \Phi^{l(t)}=\mathrm{diag}\!\big(e^{j\theta^{l(t)}_{1}},\ldots,e^{j\theta^{l(t)}_{M}}\big)$.
  \STATE Compute instantaneous SINR $\gamma_{k,\ell,n}^{(t)}$ on all active triplets {{$(k,\ell,n)\in\mathcal{A}(\mathbf{Z})$}} via \eqref{eq:SINR_pwr}.
  \STATE Find worst-link $(k^*,\!\ell^*,\!n^*)$ and set $\mathcal L^{(t)}$ via \eqref{loss-fun}.
  \STATE Compute $\mathcal{G}^{(t)}=\left\{-\frac{\partial \gamma_{k^*,\ell^*,n^*}^{(t)}}{\partial \theta_m^{l(t)}}\right\}_{l,m}$ via \eqref{Grad}, and broadcast $\mathcal{G}^{(t)}$ to all $L$ metasurface layers.
  \STATE \textit{// Layer-wise projected update with shared step $\eta^{(t)}$}
  \FOR{$l=1$ \TO $L$}
    \FOR{$m=1$ \TO $M$}
      \STATE Update phase shift $\theta_{m}^{l(t+1)} \leftarrow {\text{ Proj}}_{[0,2\pi)}\big(\theta_{m}^{l(t)} -\,\eta^{(t)} \cdot \mathcal{G}^{(t)}(l,m)\big)$.
    \ENDFOR
  \ENDFOR
\ENDFOR
\STATE Final loss over $\mathcal{A}(\mathbf{Z})$ for training via \eqref{lf}.
\FOR{$t=0$ \TO $T-1$}
\STATE \textit{// Backpropagate $\mathcal L_{\text{total}}$ through the unfolded PGD}
  \STATE Compute $\frac{\partial \mathcal L_{\text{total}}}{\partial \eta^{(t)}}$ by chain rule \eqref{backpropa}. 
  \STATE Update step size $\eta^{(t)} \leftarrow \text{Adam}\big(\eta^{(t)}, \frac{\partial \mathcal L_{\text{total}}}{\partial \eta^{(t)}}\big)$.
\ENDFOR
\ENDFOR
\STATE \textbf{Output:} Learned step-size schedule $\{\eta^{(t)}\}$ and final phases $\boldsymbol{\Theta}^{(T)}$.
\end{algorithmic}
\end{algorithm}

\begin{remark}
\textnormal{Once trained, the proposed UPGD-Net acts as a generalized SINR-aware SIM phase-shift predictor: given any new bitstream with a derived $\mathbf{Z}$, the network instantaneously outputs an optimized phase solution without requiring iterative solving. Our UPGD-Net embodies a hybrid design: 
\begin{enumerate}
\item \textbf{Model-driven structure:} The architecture strictly follows the iterative PGD structure with monotone descent, leveraging domain knowledge from fully-analog beamforming and wave-based SIM optimization.
\item \textbf{Data-driven learning:} The step sizes and update schedule are learned from data, adapting to practical channel variations and transmission scenarios.
\end{enumerate}
}
\end{remark}

\subsection{Complexity Analysis}
~~~For each subcarrier $n$ and iteration~$t$, the proposed UPGD-Net executes three groups of operations:
\emph{(i) SIM‐cascade construction.}
Because every $\mathbf\Phi^{\ell(t)}$ is diagonal,
forming the cascade $\mathbf G_{n}^{(t)}$ amounts to $L$ diagonal–matrix and dense–matrix products, costing $\mathcal O(LM^{2})$ real flops per subcarrier.
\emph{(ii) SINR calculation.}
Computing $K$ useful-signal powers and $K(K\!-\!1)$ interference terms requires $\mathcal O(K^{2}M)$ flops per subcarrier. Besides, calculating the partial derivative of the SINR w.r.t. each SIM phase shift $\theta_m^\ell$ involves a SIM-Jacobian update of $\mathcal O(M^2)$ and numerator/denominator derivatives of $\mathcal O(K^{2}M)$.
\emph{(iii) Phase update.}
Element-wise scaling by $-\eta^{(t)}\,\partial\gamma_{k,\ell,n}^{(t)}/\partial\theta_{m}^{\ell(t)}$ and $2\pi$ projection cost $\mathcal O(LM)$ flops and is negligible compared with the previous terms.

The gradients with respect to all \(LM\) phase variables are computed jointly via reverse-mode differentiation through the SIM cascade, rather than by recomputing a separate Jacobian for each phase variable. Hence, the gradient-evaluation cost remains of the same order as one backward pass through the cascade, i.e., comparable to the forward cost up to a constant factor.  Therefore, the per-iteration forward complexity is of order $\mathcal C_{\text{fwd}}
       =\mathcal O\bigl(
       N_{c}\,( L\,M^{2}\,+\,K^{2}\,M )
       \bigr)$.

During training, back-propagation reuses all forward intermediates and adds, per stage, only one vector inner product and a scalar update for the trainable step size $\eta^{(t)}$.  Hence
$\mathcal C_{\mathrm{bwd}}\approx\mathcal C_{\mathrm{fwd}}$,
and the total complexity of UPGD-Net is approximated as $\mathcal O\bigl(
       N_{c}\,T\,( L\,M^{2}\,+\,K^{2}\,M )       \bigr)$.

\section{Results and Analysis}
\label{Section5}
~~~The numerical results are presented in this section to characterize the performance of the proposed SIM-enhanced multiuser OFDM-IM communication system.
\subsection{Simulation Configurations}
~~~The system operates at a center frequency $f_0 \!=\! 28$ GHz over a bandwidth $B_w \!=\! 60$ MHz, and the number of subcarriers is set as $N_c \!=\! 16$. The BS is mounted at a height of $10$ m, while the $K$ single-antenna UEs are placed $250$ m away on a line-of-sight street, spaced $30$ m apart. Following 3GPP TR 38.901 guidelines for mmWave outdoor micro-cells~\cite{3GPP}, we set $P_k\!=\!10$ dominant taps with delays. The CP length is $N_{cp}$ = 8 samples, where $T_s = 1/B_w$ denotes the sampling interval. The condition $N_{cp} T_s$ is greater than the maximum delay $\tau_{p,\max}$, thereby mitigating inter symbol interference. The thickness of SIM is set as $D_m\!=\!0.05$ m, and the spacing between the layers is set as $d_m \!=\! D_m/L$. The parameters of SIM are configured with $S \!=\! K\!=\!4$, $M \!=\! 100$, $L\!=\! 7$, $r_{m}= c/(2 f_0)$, and $S_{m}\!=\! c^2/(2f_0)^2$. The transmit power $P_t$ is set to 10~dBm and the antenna gains of the BS and UEs are set to $5$ dBi and $0$~dBi, respectively. The noise power spectral density is -174~dBm/Hz. The BS power allocation matrix $\mathbf p$ is optimized based on the iterative water-filling algorithm~\cite{Wideband-SIM2}. {{For active subcarrier index selection, $L_b \!=\! 4$, $N \!=\! 4$, and $V \!=\! 2$ are firstly configured in TABLE~\ref{selection}, where $q_1 \!=\! 2$ bits is used to determine the indices of the two active subcarriers out of four subcarriers.}}

\begin{table}[t]
\centering
\caption{\centering{\protect\\{\textsc{Index selection table for $N = 4$, $V = 2$, and $q_1 = 2$.}}}}\
\label{selection} 
	\setlength{\tabcolsep}{0.8pt} 
	\renewcommand\arraystretch{1.3} 
\newcommand\xrowht[2][0]{\addstackgap[.5\dimexpr#2\relax]{\vphantom{#1}}}
\label{ofdmss}
\begin{tabular}{|m{2cm}<{\centering}|m{2cm}<{\centering}|m{4cm}<{\centering}|}
\hline
{\textbf{Bits}} & {\textbf{Indices} ${(\mathbf{I}_{s,\ell})}^T$} & \textbf{OFDM-IM subblocks} ${(\mathbf{x}_{s,\ell})}^T$  \\ \hline 
[0, 0] &[1, 3]  &$[\mathbf{x}_{s,\ell,1},~0,~\mathbf{x}_{s,\ell,2},~0]$ \\ \hline 
[0, 1] &[2, 4]  &$[0,~\mathbf{x}_{s,\ell,1},~0,~\mathbf{x}_{s,\ell,2}]$ \\ \hline 
[1, 0] &[1, 4]  &$[\mathbf{x}_{s,\ell,1},~0,~0,~\mathbf{x}_{s,\ell,2}]$ \\ \hline 
[1, 1] &[2, 3]  &$[0,~\mathbf{x}_{s,\ell,1},~\mathbf{x}_{s,\ell,2},~0]$ \\ \hline 
\end{tabular}
\vspace{-0.5 cm}
\end{table}

UPGD-Net implements the physical-algorithmic unrolling in Fig.~\ref{UPGD}: each column corresponds to one projected-gradient $\boldsymbol\theta$-update block. The step-size schedules $\{\eta^{(t)}\}_{t=0}^{T-1}$ constitute the only trainable parameters with $T\!=\!30$ unfolded stages. {{The learning database comprises 5,000 pairs of independently input tuples $\mathcal{H}$ for different bitstream data, while an 80\%:20\% split is used for training and validation.}} Offline optimization relies on the Adam algorithm with a learning rate of $1\times 10^{-3}$, a batch size of 64, and 120 training epochs.

\subsection{Convergence of the Proposed UPGD-Net}
~~~UPGD-Net is benchmarked against (i) conventional PGD with a fixed step $0.15$ and $I\!=\!50$ iterations and (ii) CVX solving the mathematical  reformulation with per-frame restart. All methods employ the same water-filling power allocation strategy, SIM budget, and simulation parameters.

\begin{figure}
	\centerline{\includegraphics[width=0.4\textwidth]{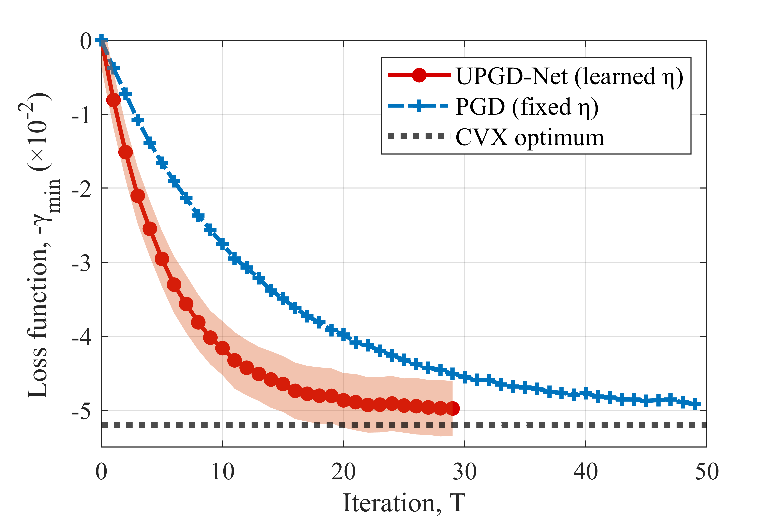}}
	\caption{{Convergence performance of the worst-link loss $\mathcal{L} = - \gamma_{\rm{min}}$ versus the iteration index $T$.}}
	\label{lossfun} 
\vspace{-0.5 cm}
\end{figure}

Fig.~\ref{lossfun} confirms the superiority of the proposed UPGD-Net design. The solid red curve is the mean performance of the proposed UPGD-Net with learned step sizes, averaged over 1,000 independent phase-shift realizations and randomly generated OFDM-IM activation matrices $\mathbf{Z}$. The surrounding shaded area shows the corresponding $\pm 1~\!\sigma$ range, i.e., the region bounded by the 16-th and 84-th percentiles of the 1,000 trajectories, illustrating the spread induced by bitstream variability. The dashed blue curve depicts conventional PGD with an a priori tuned, fixed step, {{while the dotted black line marks the CVX benchmark obtained by an offline optimization algorithm in~\cite{Wideband-SIM2}}}. Starting from the same initial condition, UPGD-Net drives the loss below -$4.5\times 10^{-2}$ within 12 unfolded stages and reaches -$5\times 10^{-2}$ after 25 stages, less than $0.2\times 10^{-2}$ away from the CVX global optimum. In contrast, the fixed-step PGD requires almost 50 iterations to attain a comparable level, demonstrating a convergence speed-up of 2.5 times. The red band further highlights the generalization capability: although every realization employs a different index bitstream, with a different subcarrier activation matrix $\mathbf{Z}$, the learned step-size schedule keeps all trajectories tightly clustered. This behavior underpins the data-driven component of the network, while the strictly PGD-based layer structure preserves the model-driven monotonic descent, ensuring that no sample ever diverges.

\begin{figure}
	\centerline{\includegraphics[width=0.4\textwidth]{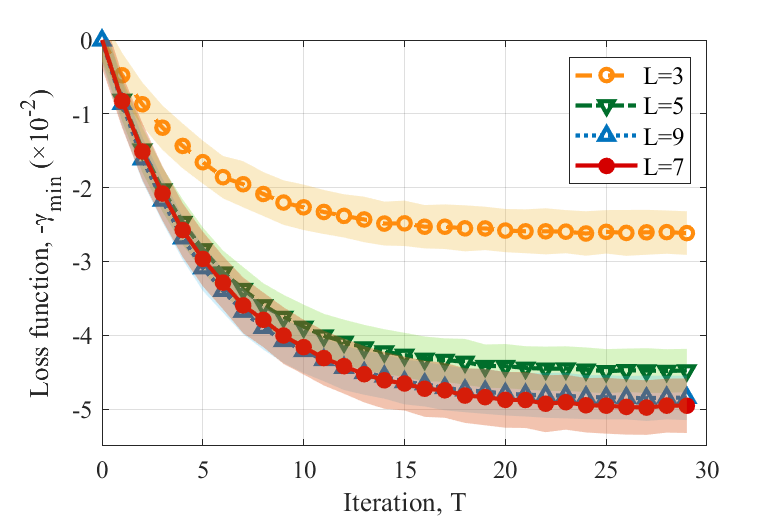}}
	\caption{{Convergence performance of the worst-link loss $\mathcal{L}$ for different numbers of SIM layers $L$ with $M=100$ meta-atoms.}}
	\label{lossfun2} 
\vspace{-0.5 cm}
\end{figure}

Fig.~\ref{lossfun2} illustrates how the number of transmit metasurface layers $L$ influences the optimization carried out by the proposed UPGD-Net. With a fixed per-layer aperture with the number of meta-atoms $M\!=\!100$, stacking more layers provides additional DoF, so the proposed UPGD-Net converges faster and reaches a lower loss from increasing $L\!=\!3$ to $L\!=\!7$. When $L$ increases further to 9, though the performance converges fast in the initial iteration, it slightly degrades in the final part. This is because neighboring layers become strongly electromagnetically coupled, so each additional layer under a fixed thickness manipulates the wavefront less effectively when exceeding a certain threshold. A larger $L$ also enlarges the two-dimensional unfolding grid, increasing per-iteration complexity, as every projected-gradient update has to account for the phase-shift tensors of all preceding layers. Under the present hardware budget, $L\!=\!7$ therefore emerges as the empirical sweet spot, striking a trade-off between worst-link SINR optimization and implementation cost. Each curve represents the ensemble mean over 1,000 random OFDM-IM activation matrices, while the surrounding shaded area marks the corresponding $\pm 1~\sigma$ interval. These results confirm that the {{coupled physical-algorithmic unrolling schedule}} on training iteration and metasurface layer generalizes well across various bit-stream variability, underscoring the robustness of the proposed model-/data-hybrid design.

\subsection{End-to-end System Performance Evaluation}
\begin{figure}
	\centerline{\includegraphics[width=0.4\textwidth]{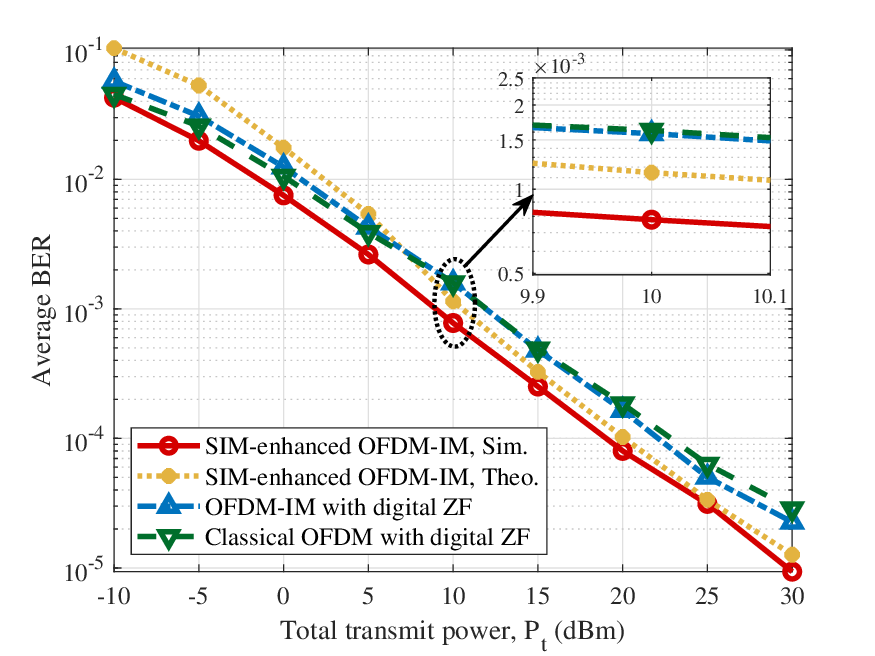}}
	\caption{{Average BER versus total transmit power for the proposed SIM-enhanced OFDM-IM and two baselines: 
digital-ZF OFDM-IM and digital-ZF full-tone OFDM.}}
	\label{fig:ber_comp} 
\vspace{-0.5 cm}
\end{figure}

~~~Fig.~\ref{fig:ber_comp} depicts four average BER curves versus the total transmit power $P_t$ on the proposed $4\!\times\!4$ SIM-enhanced multiuser MIMO downlink. All schemes are normalized to the same spectral efficiency $\eta\!=\!2.67$ bit/s/Hz using binary phase shift keying (BPSK) modulation. The results of the proposed SIM-enhanced OFDM-IM are shown both as a Monte Carlo simulation in a red solid line with circles and as the theoretical upper bound in an orange dotted line with diamonds. As the power increases, the theoretical curve becomes very tight with the simulation curve, confirming the accuracy of the pairwise-error probability analysis in Section~\ref{Section3}. In addition, two baselines are added for reference, where OFDM-IM with digital ZF is shown in blue dash-dot triangles and classical OFDM with digital ZF is in green dashed inverted triangles. Compared with normal OFDM-IM with digital ZF, the SIM-enhanced OFDM-IM architecture enjoys a consistent $2\,$--$\,3$ dB advantage at the target $ \text {BER}\!=\!10^{-3}$ owing to the wave-domain gain delivered by the large-aperture SIM structure. Both OFDM-IM curves outperform classical OFDM at medium and high powers because the index modulation exploits the same bandwidth while concentrating energy on the $V\!=\!2$ active tones, thereby enlarging the minimum Euclidean distance. Overall, this figure verifies that the SIM-enhanced OFDM-IM architecture simultaneously leverages wave-domain aperture gain and index modulation diversity to achieve the lowest BER for a given spectral efficiency and processing budget.

\begin{figure}
	\centerline{\includegraphics[width=0.4\textwidth]{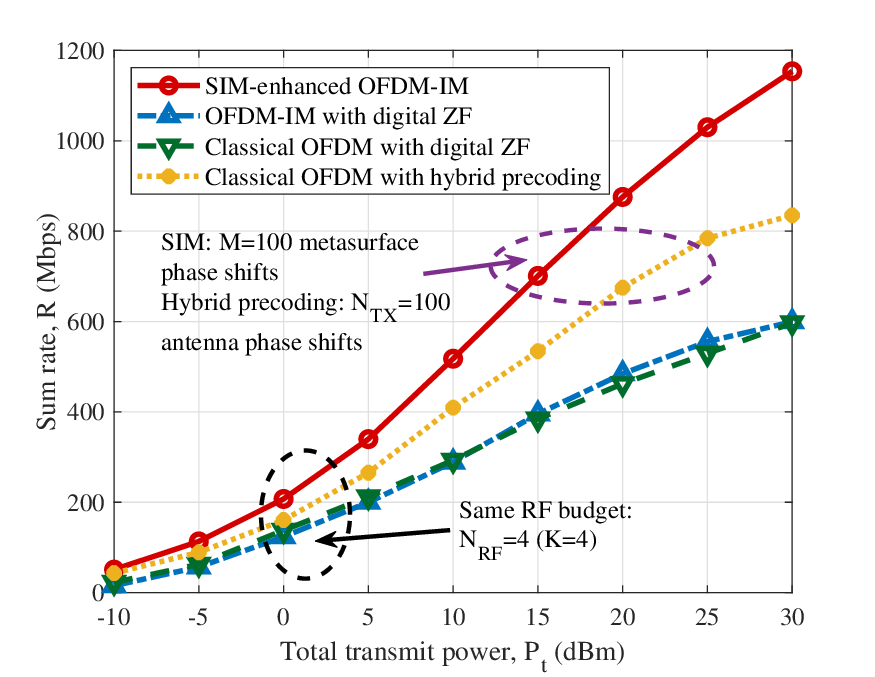}}
	\caption{{{{Sum rate versus total transmit power for the proposed SIM-enhanced OFDM-IM and three baselines: digital-ZF OFDM-IM, digital-ZF full-tone OFDM, and hybrid-precoding OFDM. }}}}
	\label{fig:sr_comp} 
\vspace{-0.5 cm}
\end{figure}

\begin{figure}
	\centerline{\includegraphics[width=0.4\textwidth]{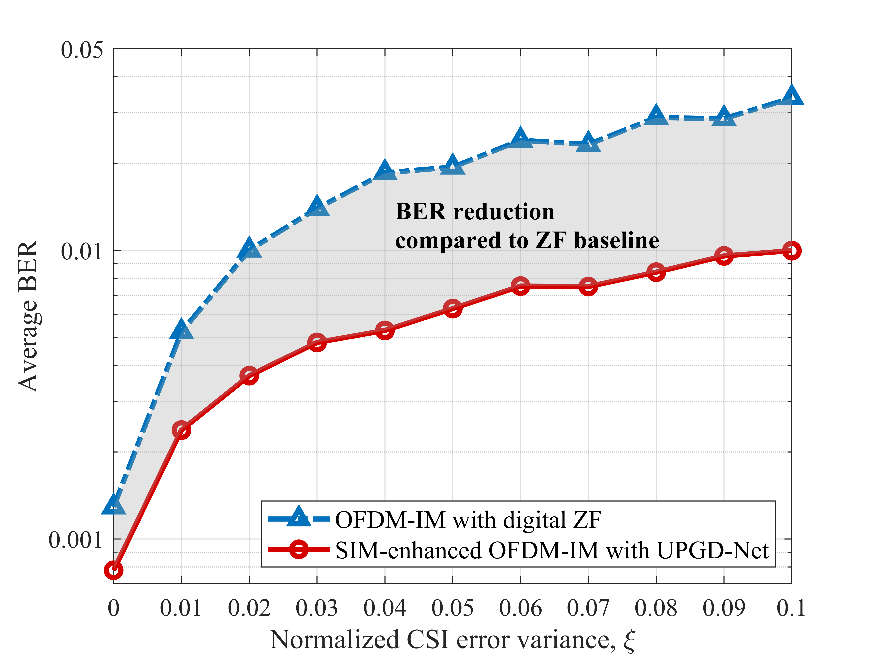}}
	\caption{{Average BER versus normalized CSI error variance for the proposed SIM-enhanced OFDM-IM structure with UPGD-Net and the digital ZF baseline.}}
	\label{fig:ber_csi} 
\vspace{-0.5 cm}
\end{figure}

{{Fig.~\ref{fig:sr_comp} depicts the sum rate versus the total transmit power for four schemes: the proposed SIM-enhanced OFDM-IM, a digital-ZF OFDM-IM baseline, a classical full-tone OFDM with digital ZF, and a large-aperture hybrid precoding OFDM baseline. To ensure a fair comparison in terms of active hardware complexity, all schemes are constrained to the same RF budget of $N_{RF}\!=\!4$ (serving $K\!=\!4$ users). Across the entire transmit power sweep, the proposed SIM-enhanced OFDM-IM curve lies $40\,$--$\,80\%$ above its digital ZF counterparts. The gain is attributed to the large-aperture SIM ($M\!=\!100$ controllable metasurface phase shifts per layer), which adds a wave-domain boost to every active subcarrier. Unlike digital ZF, which often suffers from a severe power normalization penalty to satisfy the total power constraint, the SIM redistributes signal energy more effectively via diffractive propagation. It is also worth noting that the two digital-ZF baselines exhibit closely matched sum-rate curves. This is because, under the BPSK modulation scheme, the spectral efficiency loss of IM is compensated by the power concentration on active subcarriers, resulting in a comparable overall payload to that of classical full-tone OFDM under the same system budget.

In Fig.~\ref{fig:sr_comp}, we further include a hybrid precoding OFDM benchmark that uses a conventional fully connected phase-shifter network to drive a 100-element antenna array with the same 4 RF chains. As expected, the larger antenna aperture improves the achievable sum rate over the $N_{TX}=4$ digital baselines, especially in the low-SNR regime. Nevertheless, the proposed SIM-enhanced OFDM-IM still offers a substantial gain, highlighting that the multi-layer structure can exploit a large passive aperture more effectively than a conventional hybrid precoder under the same RF-chain constraint. In practice, the hybrid precoder relies on an active aperture requiring 100 power amplifiers and hundreds of active phase shifters to drive the antenna array, while the SIM achieves equivalent performance using a quasi-passive multilayer structure with only 4 active RF chains. Consequently, the proposed SIM architecture offers a decisive advantage in terms of power consumption and hardware cost compared to conventional active phased arrays. This benefit is even more significant when compared against a fully-digital massive-MIMO benchmark, which would necessitate scaling the number of RF chains linearly with the number of antennas (i.e., $N_{RF}\!=\!N_{TX}\!=\!100$).

To address practical concerns regarding imperfect CSI at the BS, we evaluate the robustness of the proposed BER-driven UPGD-Net design. Fig.~\ref{fig:ber_csi} depicts the average BER performance versus the normalized CSI error variance, $\xi \in [0, 0.1]$, for both the proposed SIM-enhanced OFDM-IM scheme and the digital ZF baseline. As expected, the BER for both schemes degrades with increasing CSI error due to the mismatch between the true and estimated channels. However, the degradation trends differ significantly. The digital ZF baseline exhibits a rapid performance deterioration as the error variance grows. This behavior stems from the inherent sensitivity of ZF precoding to channel estimation errors, where matrix inversion amplifies estimation noise, resulting in severe residual inter-user interference. In contrast, the proposed SIM-based approach demonstrates superior robustness with a much more graceful BER degradation, thereby widening the performance gap over the ZF baseline at moderate-to-high error levels. This resilience is attributed to the gradient-based wave-domain optimization inherent in the SIM architecture and the unfolded UPGD-Net. The UPGD-Net leverages gradient-based wave-domain optimization to enhance the effective SINR margin in a BER-driven manner. Consequently, the learned beam pattern maintains its effectiveness even under CSI perturbations, validating the practical applicability of the proposed design in realistic TDD deployments.}}

\subsection{Trade-off of the Index Selection in OFDM-IM}
\begin{figure}
	\centerline{\includegraphics[width=0.4\textwidth]{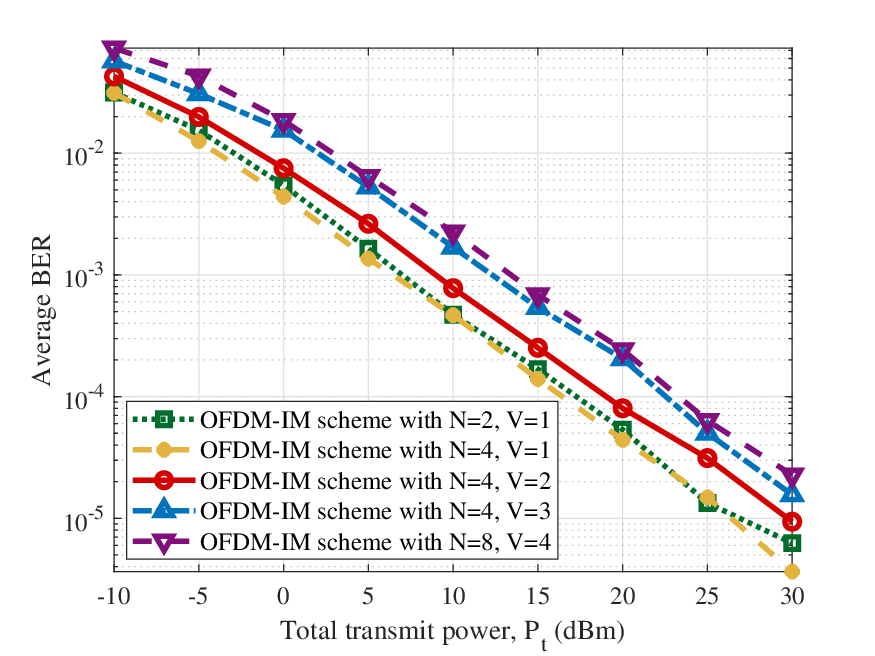}}
	\caption{{Average BER versus the total transmit power $P_t$ for five OFDM-IM configurations with different N and V values.}}
	\label{fig:ber_pt} 
\vspace{-0.3 cm}
\end{figure}

~~~Fig.~\ref{fig:ber_pt} compares the average BER of five activation patterns to analyze the trade-off of index selection. All curves retain the BPSK $1/P_t$ slope expected under SIM-enhanced wave-domain ZF, sweeping the total transmit power from $-10$ to $30$ dBm. {{TABLE~\ref{tab:se_complexity} summarizes the spectral efficiency of MIMO OFDM-IM $\eta_s = (K\,q\,L_b )/(N_c+N_{cp})$ and the per-OFDM symbol detector size $L_b 2^{q_1}M_s^V$. For the BPSK case considered here, the ML detector enumerates $2^{q_1+V}$ candidates per block.}}

\begin{table}[t]
\centering
\caption{\centering{\protect\\{\textsc{Throughput and ML-Detection Complexity per OFDM-IM Pattern under the BPSK Setting.}}}}\
\label{tab:se_complexity}
	\setlength{\tabcolsep}{0.8pt} 
	\renewcommand\arraystretch{1.3} 
\newcommand\xrowht[2][0]{\addstackgap[.5\dimexpr#2\relax]{\vphantom{#1}}}
\label{ofdmss}
\begin{tabular}{|m{2.3cm}<{\centering}|m{1.75cm}<{\centering}|m{1cm}<{\centering}|m{2.8cm}<{\centering}|}
\hline
\textbf{Pattern} & \textbf{$\eta$, bit/s/Hz}& \textbf{$L_b$} & \textbf{Candidates / OFDM} \\ \hline 
$N=2$, $V=1$ & $2.67$ & $8$ & $8\!\times\!(2\!\times\!2)\!=\!32$ \\ \hline 
$N=4$, $V=1$ & $2$ & $4$ & $4\!\times\!(4\!\times\!2)\!=\!32$ \\ \hline 
$N=4$, $V=2$ & $2.67$ & $4$ & ${{4\!\times\!(4\!\times\!4)\!=\!64}}$ \\ \hline 
$N=4$, $V=3$ & $3.33$ & $4$ & $4\!\times\!(4\!\times\!8)\!=\!128$ \\ \hline 
$N=8$, $V=4$ & $3.33$ & $2$ & ${{2\!\times\!(64\!\times\!16)\!=\!2048}}$ \\ \hline 
\end{tabular}
\vspace{-0.5 cm}
\end{table}

When considering equal spectral efficiency, i.e., $\eta_s\!=\!2.67$ bit/s/Hz, the performance of eight $2$-subcarrier subblocks with $V\!=\!1$ outperforms $N\!=~\!4$, $V\!=\!2$ by $1.2$ dB at $ \text {BER}=10^{-3}$, because every active subcarrier enjoys the full symbol power $P_t/V$. However, it also introduces twice as many subblocks, increasing the index-mapping and ML search overhead linearly with $L_b$. Raising the throughput to $\eta_s\!=\!3.33$ bit/s/Hz using scheme $N\!=\!4$, $V=3$ or scheme $N\!=\!8$, $V\!=\!4$ incurs a $\sim$3 dB penalty. This is because increasing the active subcarrier from 2 to 3 will further increase the complexity of SIM phase-shift optimization with high-dimensional activation matrix $\mathbf{Z}$. Moreover, the latter scheme further explodes the ML search space, demonstrating that merely enlarging the subblock while increasing $V$ offers no spectral-efficiency gain over the original scheme but markedly higher complexity. Finally, the low-rate scheme $N\!=\!4$, $V\!=\!1$ with $\eta_s\!=\!2$ bit/s/Hz confirms that BER performance is monotonic in $V$ for fixed symbol power, while the number of subblocks $L_b$ affects only implementation overhead. These results highlight a three-way trade-off: reliability favors small-$V$/small-$N$ patterns ($N\!=\!2$, $V\!=\!1$) but more OFDM-IM modules; throughput favors moderate-$V$ choices ($N\!=\!4$, $V\!=\!2$) with considerate reliability; and computational practicality disfavors large-subblock, large-$V$ configurations ($N\!=\!8$, $V\!=\!4$).  The subsequent simulations therefore focus on the moderate $N=4$, $V=2$ operating point, which balances BER, spectral efficiency, and detector complexity under the perfect CSI assumption adopted in the simulations.

\subsection{PAPR Reduction and RF Efficiency Analysis}
\begin{figure}[t]
    \centering
    \subfigure[\label{fig:subfig1}]{\includegraphics[width=0.5\textwidth]{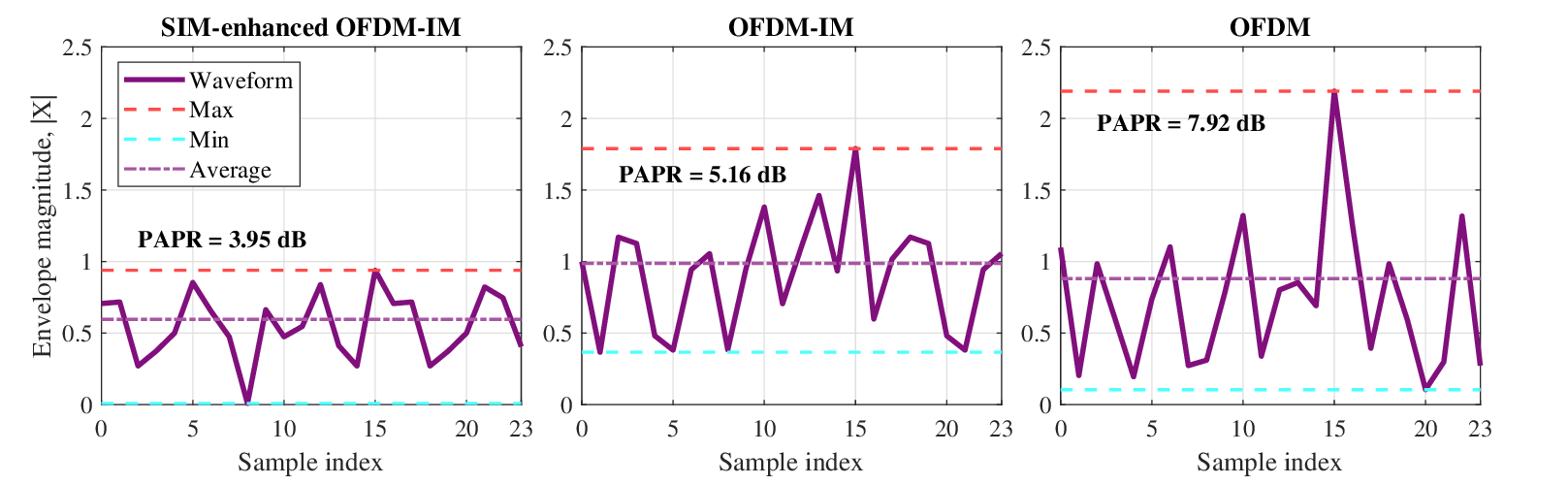}}
    \subfigure[ \label{fig:subfig2}]{\includegraphics[width=0.4\textwidth]{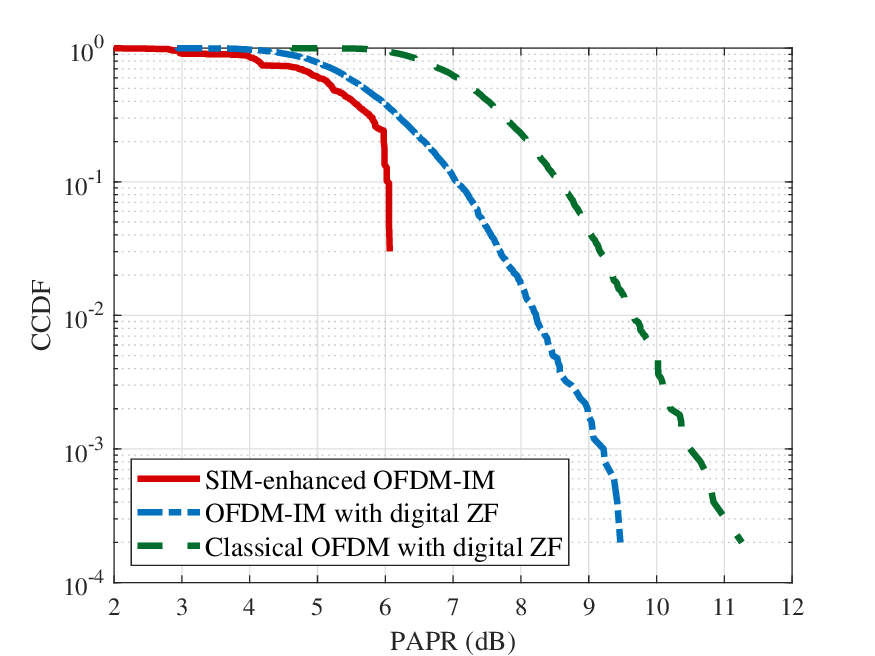}}
\caption{{{PAPR characteristics and time-domain envelope analysis: (a) Instantaneous envelope magnitude $|x|$ at the TX antenna. (b) Statistical CCDF of the PAPR over randomly generated OFDM symbols and all TX antennas.}}}
    \label{PAPR}
\vspace{-0.3 cm}
\end{figure}

~~~{{Finally, Fig.~\ref{PAPR} characterizes the PAPR of the three considered schemes: (i) SIM–enhanced OFDM–IM with wave–domain ZF, (ii) OFDM–IM with conventional digital ZF, and (iii) classical full–subcarrier OFDM with digital ZF. Fig.~\ref{fig:subfig1} first illustrates the instantaneous envelope $|x(t)|$ at a representative TX antenna for one OFDM symbol. The dynamic range between the maximum (red dashed) and average (cyan dashed) amplitudes directly reflects the PAPR level. The SIM–enhanced OFDM–IM waveform exhibits the flattest envelope, whereas OFDM–IM with digital ZF shows moderate peaks, and conventional OFDM with digital ZF exhibits the most severe fluctuations.

To provide a statistical assessment, Fig.~\ref{fig:subfig2} plots the complementary cumulative distribution function (CCDF) of the PAPR, evaluated over $10^{4}$ randomly generated OFDM symbols and all four TX antennas, with all schemes normalized to the same average transmit power. The proposed SIM-enhanced OFDM-IM clearly achieves the most favorable PAPR statistics: its CCDF curve lies consistently to the left of the two digital baselines, and at a CCDF level of $10^{-3}$, the required back-off is approximately $6$ dB, compared with about $8$--$9$~dB for OFDM–IM with digital ZF and roughly $10.5$--$11$~dB for classical OFDM with digital ZF. This gain originates from two mechanisms: First, beamforming and multiuser interference suppression are realized in the wave domain by the SIM, so each power amplifier amplifies a single-stream OFDM-IM waveform rather than a digitally superimposed multi-stream signal, thereby avoiding the mixing-induced PAPR penalty of conventional MIMO precoding. Second, index modulation activates only a subset of subcarriers in each subblock, which reduces the instantaneous power variance relative to full-tone OFDM. Overall, the observed $2$--$5$~dB PAPR reduction of the SIM-enhanced architecture over digital ZF baselines directly translates into more relaxed power-amplifier back-off requirements and improved energy efficiency in practical implementations.}}

\vspace{-0.2 cm}
\section{Conclusions}
\label{Section6}
~~~This paper proposed a SIM-enhanced multiuser OFDM-IM transceiver together with a physics-guided DUN tailored for wideband communications. On the system side, OFDM-IM is integrated as a structural complement to SIM: activating only a subset of tones widens the effective equalization bandwidth attainable by a practical $L$-layer SIM while simultaneously lowering the RF-output PAPR. {{On the objective side, we establish an analytical bridge from the worst-link BER to an active-link per-tone SINR, thereby converting reliability optimization into a robust max-min SINR surrogate that enables phase optimization under dynamic index-power constraints.}} On the solver side, we unfold PGD into the proposed UPGD-Net, which enforces unit-modulus feasibility and monotone descent layer-wise and learns only a small set of per-iteration step sizes to support frame-rate reconfiguration under varying activation matrices.

Comprehensive simulations have corroborated these design choices. The proposed framework converges rapidly and stably, typically requiring $2\,$--$\,3\times$ fewer iterations than fixed-step PGD for comparable accuracy and exhibiting a clear layer-depth sweet spot. Compared with conventional OFDM-IM with digital ZF, the SIM-enhanced OFDM-IM achieves a consistent $2\,$--$\,3$ dB advantage at the target $\text {BER}\!=\!10^{-3}$. {{Crucially, its sum rate curve surpasses that of a large-aperture hybrid benchmark with $100$ antennas in the high-SNR regime, while offering decisive energy efficiency benefits due to the quasi-passive SIM structure. Furthermore, the proposed BER-driven design demonstrates strong resilience to channel estimation errors, maintaining effective beamforming where conventional zero-forcing fails. An activation–layout sweep further shows that the operating point $(N,V)\!=\!(4,2)$ strikes a favorable balance among BER, spectral efficiency, and detector complexity under perfect CSI.}} Finally, time-domain waveform analysis confirms a reduced $2\,$--$\,5$ dB PAPR, underscoring the RF efficiency of the proposed architecture.

Beyond validating feasibility, this paper yields two design insights. First, structure design can broaden the bandwidth of analog processors: frequency-domain sparsity from OFDM-IM expands the usable equalization bandwidth of a finite-depth SIM. Second, the model-based network gains enhanced understanding capability of data: embedding exact PGD dynamics in a learnable scaffold delivers near-optimal performance at frame-level latency while preserving interpretability and constraint satisfaction. These results highlight a practical path to reliability-aware, hardware-efficient wideband multiuser MIMO OFDM systems based on programmable metasurfaces.

\appendices
\section{Proof of Lemma 1}\label{app:grad}
~~~For the $m$-th meta-atom in the $l$-th layer, define the selection matrix $\mathbf E_m=\operatorname{diag}(\mathbf e_m),
$
where $\mathbf e_m\in\mathbb R^{M}$ is the $m$-th standard basis vector. Since $\mathbf\Phi^{l}=\operatorname{diag}\!\big(e^{j\theta_1^{l}},\dots,e^{j\theta_M^{l}}\big)$, its derivative with respect to $\theta_m^{l}$ is $   \frac{\partial\mathbf\Phi^{l}}{\partial\theta_m^{l}}
     = j\,\mathbf\Phi^{l}\mathbf E_m.$ 

Split the chain around layer $l$: $\mathbf G_n   = \mathbf G_n^{(>l)}\,\mathbf\Phi^{l}\mathbf W_n^{l}\,     \mathbf G_n^{(<l)}$, where $\mathbf G_n^{(>l)}\!=\!\prod_{\nu=L}^{l+1}\mathbf\Phi^{\nu}\,\mathbf W_n^{\nu}$ and $\mathbf G_n^{(<l)}\!=\!\prod_{\nu=l-1}^{1}\mathbf\Phi^{\nu}\, \mathbf W_n^{\nu}$. Taking the derivative, we arrive at
\begin{equation}
  \frac{\partial\mathbf G_n}{\partial\theta_m^{l}}
    = j\,\mathbf G_n^{(>l)}\,
	   \mathbf\Phi^{l}\,
        \mathbf E_m\,
        \mathbf W_n^{l}\,
        \mathbf G_n^{(<l)}.
\label{eq:dGn_appendix}
\end{equation}

{{Let $\mathbf e_k\in\mathbb R^K$ denote the $k$-th standard basis vector. Since $\mathbf g_{k,\ell,n}$ is the $k$-th column of $\mathbf G_n$, we have $\mathbf g_{k,\ell,n}=\mathbf G_n\mathbf e_k$, and therefore
\begin{equation}
\frac{\partial \mathbf g_{k,\ell,n}}{\partial\theta_m^{l}}
=
\frac{\partial\mathbf G_n}{\partial\theta_m^{l}}\mathbf e_k.
\label{eq:dg_appendix}
\end{equation}

Recall the $k$-th effective beamforming coefficient $a_{k,\ell,n}$ associated with subcarrier $n$ and block $\ell$ and the interference term $b_{k,j,\ell,n}$ in Lemma \ref{lemma1}. The useful-signal numerator of SINR is 
\begin{equation}
\label{dW}
   \frac{\partial W_{k,\ell,n}}{\partial\theta_m^{l}}
     = Z_{k,\ell,n}\,p_{k,\ell,n}\,
       \frac{\partial|a_{k,\ell,n}|^{2}}{\partial\theta_m^{l}},
\end{equation}
where $
   \frac{\partial|a_{k,\ell,n}|^{2}}{\partial\theta_m^{l}}
     = 2\,\Re\!\Bigl\{
         a_{k,\ell,n}^{*}\,
         \mathbf h_{k,\ell,n}\,
         \frac{\partial\mathbf g_{k,\ell,n}}{\partial\theta_m^{l}}
       \Bigr\}.
$

Then the interference-plus-noise denominator of SINR is 
\begin{equation}
\label{dU}
   \frac{\partial U_{k,\ell,n}}{\partial\theta_m^{l}}
     = \sum_{j\neq k}
        Z_{j,\ell,n}p_{j,\ell,n}
        2\,\Re\!\Bigl\{
          b_{k,j,\ell,n}^{*}\,
          \mathbf h_{k,\ell,n}\,
          \frac{\partial\mathbf g_{j,\ell,n}}{\partial\theta_m^{l}}
        \Bigr\}.
\end{equation}

Using the quotient rule,
  \begin{equation}
\label{dF}
  \frac{\partial \gamma_{k,\ell,n}}{\partial \theta_m^l} = \frac{1}{U_{k,\ell,n}^2} \left( U_{k,\ell,n} \frac{\partial W_{k,\ell,n}}{\partial \theta_m^l} - W_{k,\ell,n} \frac{\partial U_{k,\ell,n}}{\partial \theta_m^l} \right).
  \end{equation}}}

Finally, substituting \eqref{eq:dGn_appendix}--\eqref{dU} into \eqref{dF} yields the closed-form gradient used in the PGD update.

\end{document}